\documentclass[10pt]{iopart}

\usepackage{iopams}  
\usepackage{graphicx}
\usepackage{bm}
\usepackage{makecell}

\usepackage{cite}

\begin{document}

\title{A systematic study of the dynamics of chain formation in electrorheological fluids}

\author{D\'{a}vid Fertig$^{1}$, Dezs\H{o} Boda$^{1}$ and Istv\'{a}n Szalai$^{2,3}$}

\address{$^{1}$Center for Natural Sciences, University of Pannonia, Egyetem u. 10, Veszpr\'{e}m, 8200, Hungary}
\address{$^{2}$Research Centre for Engineering Sciences, University of Pannonia, Egyetem u. 10, Veszpr\'{e}m, 8200, Hungary}
\address{$^{3}$Institute of Mechatronics Engineering and Research, University of Pannonia, Gasparich M\'{a}rk u. 18/A, Zalaegerszeg, 8900, Hungary}
\ead{boda@almos.uni-pannon.hu}
\vspace{10pt}
\begin{indented}
\item[]November 2020
\end{indented}

\begin{abstract}
We report a systematic study of the dynamics of chain formation in electrorheological (ER) fluids using Brownian Dynamics simulations. 
The parameters of the system such as applied electric field, polarizability, dipole moment, friction coefficient, and number density are expressed in reduced units and changed in a wide range in order to map the system's behavior as a function of them.
We define time constants obtained from bi-exponential fits to time dependence of various physical quantities such as dipolar energy, diffusion constant, and average chain length.
The smaller time constant is associated with the formation of shorter chains (pairs, triplets, and so on), while the larger time constant is associated with the formation of longer chains in the regime of those that overarch the simulation cell.
We use the approximation that the dipole moments are induced by the applied electric field only, as usual in the literature.
However, we report preliminary results for the case when particle-particle polarization is also possible.  
\end{abstract}

%
%
\submitto{\JPCM}
%
%
\ioptwocol

\section{Introduction}

In electrorheological (ER) fluids~\cite{winslow_jap_1949} fine non-conducting solid particles are suspended in an electrically insulating liquid with the particles having larger dielectric constant than the solvent.
Then, an applied electric field induces polarization charges at the arising dielectric boundaries that can be corresponded to effective dipoles placed in the centers of the particles.

The interactions of these dipoles lead to a structural change in the ER fluid known as the ER response.
This structural change is the aggregation of ER particles first into shorter, then into longer chains due to the fact that the head-to-tail position of two dipoles along the direction of the applied field is a minimum-energy configuration.
In the case of strong applied fields, the chains form larger clusters, for example, columnar structures.

This structural change results in changes in all major physical properties of the ER fluid.
From a practical point of view, one of the most important is viscosity.
The externally controllable, fast and reversible change in viscosity makes ER fluids a central component of various devices, such as brakes, clutches, dampers, and valves \cite{Duclos_1992,havelka_progress_ER_1995}. 
Such devices have crucial importance in various industries including the automotive industry.

Because the functioning of ER devices is based on microscopic mechanisms leading to an emergent macroscopic pattern, a number of modeling studies \cite{Klingenberg_1989,Heyes_1990,whittle_jnnfm_1990,klingenberg_langmuir_1990,Jaggi_1991,see_psj_1991,bonnecaze_jcp_1992,Toor_1993,hass_pre_1993,tao_prl_1994,tao_ijmpb_1994,baxter_drayton_jr_1996,gulley_pre_1997,jian_jap_1996,blair_jcp_1999, wang_ijes_2001,enomoto_physa_2002,climent_langmuir_2004,cao_jpcb_2006,dominguezgarcia_pre_2007} have been devoted to the investigation of the microscopic processes behind chain formation and corresponding changes in measurable physical properties.
Cluster formation has been investigated via cluster size distribution \cite{Klingenberg_1989,see_psj_1991,Toor_1993,hass_pre_1993,climent_langmuir_2004,dominguezgarcia_pre_2007}, order parameters \cite{hass_pre_1993,tao_prl_1994,tao_ijmpb_1994,baxter_drayton_jr_1996,enomoto_physa_2002}, mean square displacement and diffusion constant \cite{Klingenberg_1989,whittle_jnnfm_1990,hass_pre_1993}, pair distribution functions \cite{whittle_jnnfm_1990,hass_pre_1993}, and relaxation times \cite{Heyes_1990,Toor_1993,hass_pre_1993,cao_jpcb_2006}.
In particular, Cao et al.\ \cite{cao_jpcb_2006} identified relaxation times corresponding to various subprocesses such as initial aggregation, chain formation, and column formation. 
See et al.~\cite{see_psj_1991} described aggregation kinetics and chain formation, for two dimensional ER fluids. 
A scaling law is also presented in this work.
These processes were also investigated in the presence of shear~\cite{Heyes_1990,whittle_jnnfm_1990,bonnecaze_jcp_1992,baxter_drayton_jr_1996,cao_jpcb_2006} with the goal of computing  shear stress, various terms of viscosity, oscillatory strain, and dependence on strain rate. 

In this paper, our main interest is to study the dynamics of the formation of chains with a newly developed Brownian Dynamics  simulation package based on a novel Langevin integrator \cite{Gronbech_Jensen_mp_2013,farago_physicaA_2019,jensen_mp_2019}.
We are interested in the mechanisms by which the chains and their clusters are formed when the electric field is switched on.
We characterize this dynamics by plotting various physical properties such as energy, diffusion constant, average chain length, chain length distributions, and radial distribution functions as functions of time.

In particular, we intend to provide a systematic study over a wide range of reduced parameters and to study the effect of the various parameters on the dynamics of chain formation.
Reduced parameters make it possible to study a model without any considerations of real ER fluids.
Here, we reduce our quantities with the particle diameter, $d$, mass of the particle, $m$, and the thermal energy, $kT$, where $k$ is Boltzmann's constant and $T$ is the absolute temperature.
Reduced units are also useful because they may express relative strengths of competing effects.
The reduced dipole moment, for example, is the strength of the dipole-dipole interaction relative to $kT$, namely, it expresses the relative strength of the ordering effect of the electrostatic forces compared to the disordering effect of thermal motion.

We consider the effect of system size (number of particles), the strength of the applied electric field, polarizability, the product of these (dipole moment), friction coefficient, and the number density of the ER particles.
We try to dig into the depths of microscopic processes in order to understand what is going on at the microscopic level.

Once the time dependence of the various physical quantities is available, we average the behavior of the chains of different lengths into some aggregate behaviors characterized by two time constants obtained from fitting bi-exponential functions to various physical quantities.
We interpret and justify these time constants.
Such characteristic time constants are useful because they make it possible to relate out simulations to experimental data, where such time constants can also be obtained from fitting to experimental data.~\cite{horvath_pre_2012,horvath_pre_2015}

The papers listed above were produced with an approximate model where the dipoles on the ER particles were assumed to be induced by the applied field only, while the polarization of the particles by other particles was ignored.
Here, we also use this approximation, but we also present some preliminary results for the case when particle-particle polarization is present.
These results show that the two models produce similar dynamics if the polarizability is not too large.
If the polarizability is large, however, we expect deviations from the results presented here and in the literature.

Those investigations are referred to a later paper; here we present systematically and in detail what happens in an ER fluid on the microscopic level during the time from the moment of switching the applied field on to the moment of the chains formed.

\section{Models and methods}
\label{sec:model}

\subsection{The polarizable dielectric sphere} 

\begin{figure}[b]
	\centering
	\includegraphics[width=0.2\textwidth]{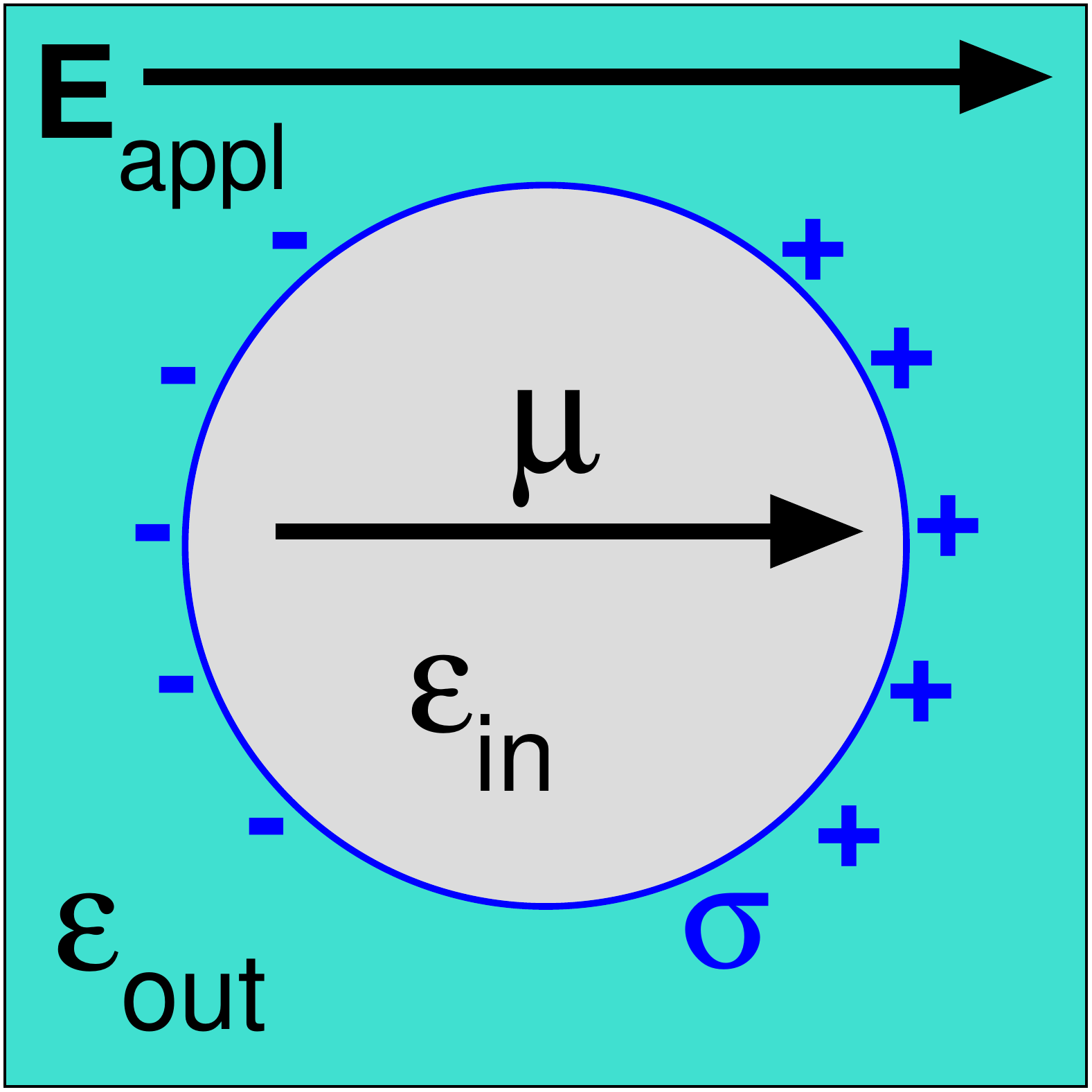}
	\caption{Sketch of an ER particle in an external electric field, $\mathbf{E}_{\mathrm{appl}}$. The dielectric constant inside the sphere is $\epsilon_{\mathrm{in}}$, while outside the sphere is $\epsilon_{\mathrm{out}}$. The surface charge distribution, $\sigma(\mathbf{r})$, induced on the dielectric boundary (Eq.~\ref{eq:sigma}) can be approximated by a point dipole, $\bm{\mu}$, in the center of the sphere (Eq.~\ref{eq:dipole}).}
	\label{fig01}
\end{figure}

The ER fluid is modeled as dielectric spheres of dielectric constant $\epsilon_{\mathrm{in}}$ inside the sphere immersed in a fluid of dielectric constant $\epsilon_{\mathrm{out}}$ (Fig.~\ref{fig01}).
The radius of the spheres is $R$, while their diameter is $d=2R$.
If an electric field, $\mathbf{E}$ is applied on the sphere, a polarization charge density 
\begin{equation}
 \sigma  (\theta) = 3 \epsilon_0 \left( \frac{\epsilon_{\mathrm{in}}-\epsilon_{\mathrm{out}}}{\epsilon_{\mathrm{in}}+2\epsilon_{\mathrm{out}}} \right) E \cos \theta,
 \label{eq:sigma}
 \end{equation} 
is induced on the surface of the sphere, where $E=|\mathbf{E}|$, $\theta$ is the polar coordinate (the angle between a point on the surface and $\mathbf{E}$), and $\epsilon_{0}$ is the permittivity of vacuum.
Far from the sphere, the effect of this surface charge distribution can be approximated with an ideal point dipole placed in the center of the sphere computed as \cite{jackson}
 \begin{equation}
  \bm{\mu}=4\pi \epsilon_{0} \left( \frac{\epsilon_{\mathrm{in}}-\epsilon_{\mathrm{out}}}{\epsilon_{\mathrm{in}}+2\epsilon_{\mathrm{out}}} \right) R^{3}\mathbf{E} = \alpha \mathbf{E}, 
  \label{eq:dipole}
 \end{equation} 
 where 
 \begin{equation}
 \alpha = 4\pi \epsilon_{0} \left( \frac{\epsilon_{\mathrm{in}}-\epsilon_{\mathrm{out}}}{\epsilon_{\mathrm{in}}+2\epsilon_{\mathrm{out}}} \right) R^{3} 
 \label{eq:alpha}
 \end{equation} 
is the particle polarizability.

If it is further assumed that the characteristic time of the rearrangement of the surface charge during the movement of the particles is much smaller than the characteristic time of the rotation of the particles, the $\bm{\mu}$ dipole always points into the direction of $\mathbf{E}$ even if the sphere rotates.
Then, the induced charges chiefly corresponding to the polarization of solvent molecules around the sphere always have enough time to rearrange themselves.
As an ideal limit, we take this rearrangement infinitely fast.

If we take a system of $N$ particles at positions $\{\mathbf{r}_{j}\}$, the potential produced by a dipole $\bm{\mu}_{j}$ (that is at $\mathbf{r}_{j}$) at the position $\mathbf{r}_{i}$ of another dipole $\bm{\mu}_{i}$ is 
\begin{equation}
 \Phi_{j}(\mathbf{r}_{i}) = \frac{1}{4\pi \epsilon_0} \frac{\bm{\mu}_{j}\cdot \mathbf{r}_{ij}}{r_{ij}^{3}}, 
\end{equation} 
where $\mathbf{r}_{ij}=\mathbf{r}_{i}-\mathbf{r}_{j}$ and $r_{ij}=|\mathbf{r}_{ij}|$.
The electric field exerted on dipole $i$ by dipole $j$ is
\begin{equation}
 \mathbf{E}_{j}(\mathbf{r}_{i}) = \frac{1}{4\pi \epsilon_0} \frac{3\mathbf{n}_{ij}(\mathbf{n}_{ij}\cdot \bm{\mu}_{j})-\bm{\mu}_{j}}{r_{ij}^{3}},
\end{equation} 
where $\mathbf{n}_{ij}=\mathbf{r}_{ij}/r_{ij}$.

In Eq.~\ref{eq:dipole}, the electric field at $\mathbf{r}_{i}$ is a sum of the applied field, $\mathbf{E}_{\mathrm{appl}}$ (it defines the $z$ direction), and the electric field produced by all the other dipoles, $\mathbf{E}(\mathbf{r}_{i})=\sum_{j\ne j} \mathbf{E}_{j}(\mathbf{r}_{i})$.
The total dipole moment 
\begin{equation}
 \bm{\mu}^{\mathrm{tot}}_{i} = \alpha \mathbf{E}_{\mathrm{appl}} + \alpha \mathbf{E}(\mathbf{r}_{i}) = \bm{\mu}^{\mathrm{appl}}_{i} + \bm{\mu}^{\mathrm{part}}_{i}
 \label{eq:iterate_mu}
\end{equation} 
is induced by these two components and is split into the terms $\bm{\mu}^{\mathrm{appl}}_{i}$ and $\bm{\mu}^{\mathrm{part}}_{i}$ accordingly.
\begin{itemize}
\item 
The dipole moment $\bm{\mu}_{i}^{\mathrm{appl}}$ induced by the applied field is constant and always points into the $z$ direction.
It can be called ``permanent'' in the sense that it is always there when $\mathbf{E}_{\mathrm{appl}}$ is switched on, but it is not permanent in the chemical sense in which the polar molecules have permanent dipoles.
\item 
The dipole moment $\bm{\mu}_{i}^{\mathrm{part}}$ induced by all the other ER particles  depends on the electric field produced by the induced dipoles, so it is calculated by an iterative procedure.~\cite{vesely_jcompp_1977} 
\end{itemize}
In the rheological literature, it is usual to ignore $\bm{\mu}^{\mathrm{part}}_{i}$, namely, the polarization of the particles by each other. 
In this work, we show some preliminary results for the full solution of Eq.~\ref{eq:iterate_mu} in comparison with the approximate approach using an effective dipole moment
\begin{equation}
\langle \bm{\mu}^{\mathrm{eff}} \rangle = \bm{\mu}^{\mathrm{appl}}  + \langle \bm{\mu}^{\mathrm{part}} \rangle ,
\end{equation} 
where $\langle \dots \rangle$ denotes ensemble average.

\subsection{Dipolar interactions between particles}

The interaction potential between the two dipoles (irrespective whether it is induced by $E_{\mathrm{appl}}$ or by other particles) is
\begin{eqnarray}
 u^{\mathrm{dip}}_{ij}(\mathbf{r}_{ij},\bm{\mu}_{i},\bm{\mu}_{j}) = - \bm{\mu}_{i}\cdot \mathbf{E}_{j}(\mathbf{r}_{i}) = \nonumber \\
= - \frac{1}{4\pi \epsilon_0} \frac{3 (\mathbf{n}_{ij}\cdot \bm{\mu}_{i}) (\mathbf{n}_{ij}\cdot \bm{\mu}_{j}) - \bm{\mu}_{i}\cdot \bm{\mu}_{j} }{r_{ij}^{3}},
\label{eq:udippair}
\end{eqnarray} 
while the force exerted on dipole $\bm{\mu}_{i}$ by dipole $\bm{\mu}_{j}$ is
\begin{eqnarray}
 \mathbf{f}^{\mathrm{dip}}_{ij}(\mathbf{r}_{ij},\bm{\mu}_{i},\bm{\mu}_{j}) = - (\bm{\mu}_{i}\cdot \nabla_{i}) \mathbf{E}_{j}(\mathbf{r}_{i}) = \nonumber \\
=  \frac{1}{4\pi \epsilon_0} 
\frac{1}{r_{ij}^{4}} \left\{ 3 \left[ 
\bm{\mu}_{i}  (\mathbf{n}_{ij}\cdot \bm{\mu}_{j}) +
\bm{\mu}_{j}  (\mathbf{n}_{ij}\cdot \bm{\mu}_{i}) + \right. \right. \nonumber \\  
+ \left. \left. \mathbf{n}_{ij} (\bm{\mu}_{i} \cdot \bm{\mu}_{j}) \right]
- 15 \mathbf{n}_{ij} (\mathbf{n}_{ij}\cdot \bm{\mu}_{i}) 
                     (\mathbf{n}_{ij}\cdot \bm{\mu}_{j}) 
\right\}  
\label{eq:fdd}.
\end{eqnarray} 
The torque acting on the dipoles has been ignored due to the assumption of the instantaneous rearrangement of induced charges.

The dipolar force acting on dipole $\bm{\mu}_{i}^{\mathrm{appl}}$ is
\begin{equation}
\mathbf{F}^{\mathrm{dip}}_{i} = \sum_{j\ne i} \mathbf{f}_{ij}^{\mathrm{appl}}  + \sum_{j\ne i} \mathbf{f}_{ij}^{\mathrm{part}},
\end{equation} 
where
\begin{equation}
 \mathbf{f}_{ij}^{\mathrm{appl}}  = \mathbf{f}^{\mathrm{dip}}_{ij}(\mathbf{r}_{ij},\bm{\mu}^{\mathrm{appl}} _{i},\bm{\mu}_{j}^{\mathrm{appl}} ) 
\end{equation} 
is the force exerted on dipole $\bm{\mu}^{\mathrm{appl}} _{i}$ by dipole $\bm{\mu}^{\mathrm{appl}} _{j}$, and
\begin{equation}
 \mathbf{f}_{ij}^{\mathrm{part}} = \mathbf{f}^{\mathrm{dip}}_{ij}(\mathbf{r}_{ij},\bm{\mu}^{\mathrm{appl}} _{i},\bm{\mu}_{j}^{\mathrm{part}}) 
\end{equation} 
is the force exerted on dipole $\bm{\mu}^{\mathrm{appl}} _{i}$ by dipole $\bm{\mu}^{\mathrm{part}} _{j}$.

Similarly, the dipolar energy is divided as
\begin{equation}
 U^{\mathrm{dip}} = U^{\mathrm{appl}}  + U^{\mathrm{part}},
\end{equation} 
where 
\begin{equation}
 U^{\mathrm{appl}}  = \frac{1}{2} \sum_{i}\sum_{j} u_{ij}^{\mathrm{dip}}(\mathbf{r}_{ij},\bm{\mu}_{i}^{\mathrm{appl}} ,\bm{\mu}_{j}^{\mathrm{appl}} )
 \label{eq:uappl}
\end{equation} 
is the interaction energy between dipoles induced by $E_{\mathrm{appl}}$, and
\begin{equation}
 U^{\mathrm{part}} = \frac{1}{2} \sum_{i}\sum_{j} u_{ij}^{\mathrm{dip}}(\mathbf{r}_{ij},\bm{\mu}_{i}^{\mathrm{appl}} ,\bm{\mu}_{j}^{\mathrm{part}})
 \label{eq:upart}
\end{equation} 
is the induction energy.
A more detailed derivation of the induction energy can be found in the paper of P\v{r}edota et al.~\cite{predota_mp_2002}

\subsection{Short-range interactions between particles}

The full interaction potential between two ER particles must contain a short-range core potential for which we use the cut \& shifted LJ potential also known as the Weeks-Chandler-Anderson (WCA) potential that is
\begin{equation}
 u_{ij}^{\mathrm{WCA}}(r_{ij}) = \left\{
 \begin{array}{ll}
  u_{ij}^{\mathrm{LJ}} (r_{ij}) + u_{ij}^{\mathrm{LJ}} (r_{\mathrm{c}}) \; & \mathrm{if} \; \; r_{ij}<r_{\mathrm{c}} \\
  0 \; & \mathrm{if} \; \; r_{ij}>r_{\mathrm{c}}
 \end{array}
\right. ,
\label{eq:ucore}
\end{equation} 
where 
\begin{equation}
 u_{ij}^{\mathrm{LJ}} (r_{ij}) = 4\varepsilon^{\mathrm{LJ}} \left[ \left( \frac{d}{r_{ij}} \right)^{12} - \left( \frac{d}{r_{ij}} \right)^{6} \right]
\end{equation} 
is the LJ potential.
The WCA force is 
\begin{equation}
 \mathbf{f}_{ij}^{\mathrm{WCA}}(\mathbf{r}_{ij}) = \left\{
 \begin{array}{ll}
  \mathbf{f}_{ij}^{\mathrm{LJ}} (\mathbf{r}_{ij})  \; &\mathrm{if} \; \; r_{ij}<r_{\mathrm{c}} \\
  0 \; &\mathrm{if} \; \; r_{ij}>r_{\mathrm{c}}
 \end{array}
\right. ,
\label{eq:fcslj}
\end{equation} 
where 
\begin{equation}
 \mathbf{f}_{ij}^{\mathrm{LJ}} (\mathbf{r}_{ij}) = 24 \varepsilon^{\mathrm{LJ}} \left[2 \left( \frac{d}{r_{ij}} \right)^{12} - \left( \frac{d}{r_{ij}} \right)^{6} \right] \frac{\mathbf{r}_{ij}}{r_{ij}^{2}}
\end{equation} 
is the LJ force.
In these equations, the cutoff distance is $r_{\mathrm{c}}=2^{1/6}d$ that is at the minimum of the LJ potential, so this potential is a smooth repulsive core potential.

\subsection{Brownian Dynamics simulation}
\label{sec:method}

The trajectories of the particles in the phase space interacting with each other via the systematic force
\begin{equation}
\mathbf{F}_{i} = \sum_{j} (\mathbf{f}_{ij}^{\mathrm{WCA}} + \mathbf{f}_{ij}^{\mathrm{appl}}  + \mathbf{f}_{ij}^{\mathrm{part}})
\end{equation} 
are determined by solving Newton's equation of motion in an MD simulation.
When the particles are immersed in a solvent, we use Langevin's equations of motion \cite{lemons_1997}
\begin{equation}
m\frac{d\mathbf{v}_{i}(t)}{dt} = \mathbf{F}_{i}\left(\mathbf{r}_{i}(t)\right) -m \gamma \mathbf{v}_{i}(t) + \mathbf{R}_{i}(t),
\label{eq:langevin}
\end{equation} 
where $\mathbf{r}_{i}$, $\mathbf{v}_{i}$, $m$, and $\gamma$ are the position, the velocity, the mass, and the friction coefficient of particle $i$, respectively.
The mass and the friction coefficient are assumed to be the same for every particle, but, in general, they can depend on $i$.

In addition to the systematic force, the force has two extra components: the frictional force, $-m \gamma \mathbf{v}_{i}(t)$, and the random force, $\mathbf{R}_{i}(t)$.
The former describes friction, while the latter describes random collisions with surrounding solvent molecules.
They represent the interactions with the heat bath and are coupled through the fluctuation--dissipation theorem.

This stochastic differential equation is solved numerically.~\cite{schneider_prb_1978,van_gunsteren_mp_1982,brunger_cpl_1984,leimkuhler_amr_2012}
We use the GJF-2GJ version~\cite{jensen_mp_2019} of a collections of algorithms proposed by Gr{\o}nbech-Jensen and Farago:~\cite{Gronbech_Jensen_mp_2013,farago_physicaA_2019,jensen_mp_2019}
\begin{equation}
 v^{n+\frac{1}{2}} = a v^{n-\frac{1}{2}} + \frac{\sqrt{b}\Delta t}{m} f^{n} + \frac{\sqrt{b}}{2m} \left( R^{n}-R^{n+1} \right)
 \label{eq:farago_v}
\end{equation}
\begin{equation}
 r^{n+1}=r^{n} + \sqrt{b} v^{n+\frac{1}{2}}  \Delta t ,
 \label{eq:farago_r}
\end{equation}
where $r^{n}=r(t^{n})$ is any position coordinate of any particle, $v^{n}=v(t_{n})$ is any velocity coordinate of any particle, $t^{n}=n\Delta t$ is the time in the $n$th time step, $\Delta t$ is the time step, 
\begin{equation}
 a = \frac{1-\gamma \Delta t/2}{1+\gamma \Delta t /2} ,
\end{equation}
\begin{equation}
 b =  \frac{1}{1+\gamma \Delta t /2} ,
\end{equation}
$t_{n+\frac{1}{2}}=t_{n}+\frac{\Delta t}{2}$, and $t_{n-\frac{1}{2}}=t_{n}-\frac{\Delta t}{2}$.
The discrete time noise  
\begin{equation}
 R^{n+1} = \int_{t_{n}}^{t_{n+1}} R(t')dt'
\end{equation}
is a random Gaussian number with properties 
\begin{equation}
 \langle R^{n} \rangle =0
\end{equation}
and
\begin{equation}
 \langle R^{m}R^{n}\rangle = 2 kT \gamma m \Delta t \delta_{mn} 
\end{equation}
with $\delta_{mn}$ being the Kronecker-delta.

\begin{table*}[t]
\caption{Reduced quantities}
\label{tab:reduced}
 	\def\arraystretch{1.3}
 	\centering
\begin{tabular}{lcll} \hline 
Quantity & Symbol & Unit quantity & Reduced quantity \\ \hline
Time & $t$  &  $t_{0}=\sqrt{md^{2}/kT}$ & $t^{*}=t\sqrt{kT/md^{2}}$ \\
Distance & $r$ & $r_{0}=d$ & $r^{*}=r/d$ \\
Density & $\rho$ & $\rho_{0}=1/d^{3}$ & $\rho^{*}=\rho d^{3}$ \\
Velocity & $v$ & $v_{0}=d/t_{0} = \sqrt{kT/m}$ & $v^{*}=v\sqrt{m/kT}$ \\
Energy & $u$ & $u_{0}=kT$ & $u^{*}=u/kT$ \\
Force & $F$ & $F_{0}=kT/d$ &  $ F^{*}=Fd/kT$ \\
Electric field & $E$ & $E_{0}=\sqrt{kT/4\pi\epsilon_0 d^{3}}$ & $E^{*}=E \sqrt{4\pi \epsilon_0 d^{3}/kT}$ \\
Dipole moment & $\mu$ & $\mu_{0}=\sqrt{4\pi\epsilon_{0}kTd^{3}}$ & $\mu^{*}=\mu/\sqrt{4\pi\epsilon_{0}kTd^{3}} $ \\
Particle polarizability & $\alpha$ & $\alpha_{0}=4\pi \epsilon_{0} d^{3}$ & $\alpha^{*} = \alpha /4\pi\epsilon_{0}d^{3}$ \\
Friction coefficient & $\gamma$ & $\gamma_{0}= \sqrt{kT/md^{2}}$ & $\gamma^{*} = \gamma \sqrt{md^{2}/kT}$ \\ \hline
\end{tabular}
\end{table*}

\subsection{Reduced units}
\label{sec:scaling}

There are competing effects in an ER system.
Since the head-to-tail position, in which the dipoles are aligned along $\mathbf{n}_{ij}$ ($\theta{=}0$) at contact ($r_{ij}{=}d$), has a minimum energy, dipolar interactions have an ordering effect. 
Thermal motion, on the other hand, has a disordering effect that expresses the coupling to a thermostat of temperature $T$ and friction with the surrounding solvent with viscosity $\eta$.
We can characterize the disordering effect of the thermal motion by $kT$ energetically. 

The balance of these competing effects can be emphasized by using reduced units in the calculations.
Reduced units express physical quantities as dimensionless numbers obtained by dividing a quantity in a physical unit by a unit quantity in the same unit, $t^{*}=t/t_{0}$, for example.
Reduced quantities are useful from a practical point of view because their values are close to $1$, so it is easier to work with them.
They are also useful because they express relations between the quantities in the numerator and the denominator.
In this work, we use the particle mass, $m$, the particle diameter, $d$, and $kT$ to build the reduced quantities (Table \ref{tab:reduced}).

Using reduced units is a kind of scaling~\cite{Heyes_1990} phenomenon, such as the theorem of corresponding states.
For a specific set of reduced parameters that define the state of the system, the system behaves in a well-defined way.
A set of reduced parameters, however, can be constructed from different sets of parameters in real-life physical units such as the temperature, $T$, the mass density of the material of the ER particle, $\rho_{\mathrm{in}}$, the diameter of the ER particle, $d$, the dielectric constant of the ER particle, $\epsilon_{\mathrm{in}}$, the dielectric constant of the solvent, $\epsilon_{\mathrm{out}}$, the viscosity of the solvent, $\eta$, and the strength of the applied electric field, $E_{\mathrm{appl}}$.

Many of these quantities enter the calculations indirectly.
The paremeters $\rho_{\mathrm{in}}$ and $d$ determine the mass, $m$, of the particle through $m=\rho_{\mathrm{in}}d^{3}\pi/6$.
The parameters $\epsilon_{\mathrm{in}}$, $\epsilon_{\mathrm{out}}$, and $d$ determine the particle polarizability, $\alpha$, through Eq.~\ref{eq:alpha}.
The friction coefficient can be computed from Stokes' law as
\begin{equation}
 \gamma =  \frac{3\pi \eta d}{m}.
\end{equation} 
The diffusion constant in the high coupling limit can be expressed by Einstein's relation:
\begin{equation}
 D=\frac{kT}{m\gamma},
 \label{eq:einsten}
\end{equation} 
or, in reduced units, $D^{*}=1/\gamma^{*}$.

An important parameter is the (square of the) reduced dipole moment:
\begin{equation}
 (\mu^{*})^{2} = \frac{\mu^{2}/4\pi \epsilon_{0}d^{3}}{kT} 
\label{eq:squaredip}
 \end{equation} 
that is related to the ratio of the dipolar energy and the thermal energy.
If $(\mu^{*})^{2}$ is large, the dipolar interactions are strong enough to induce chain formation.
If $(\mu^{*})^{2}$ is too large, the chains freeze, and the ER particles solidify (note that the fluid itself does not solidify). 
If $(\mu^{*})^{2}$ is small, thermal motion prevents chain formation and/or breaks the chains.

Our practical concern is how to make the simulation efficient enough to collect enough information about the dynamics of the system in a reasonable amount of computer time.
The time step, $\Delta t^{*}$, with which we can tune the speed of sampling is a subject of optimization.
If $\Delta t^{*}$ is too small, the simulation will evolve slowly at the price of valuable computation time.
If $\Delta t^{*}$ is too large, the overlap of the repulsive cores of the particles (Eq.~\ref{eq:fcslj}) leads to instabilities in solving the Langevin equation.
If one wants to use a large time step, there are methods to cope with this problem~\cite{whittle_jnnfm_1990,tao_prl_1994,berti_jctc_2014}. 

Without such techniques, we can give an estimation for a maximum value of $\Delta t^{*}$ above which there is a considerable risk for particle overlap.
We introduce the average distance, $\overline{\Delta s}$, that a particle moves in a time step with the average thermal velocity, $\bar{v}=\sqrt{3kT/m}$. 
In reduced units it is
\begin{equation}
 \overline{\Delta s^{*}} = \frac{\bar{v}\Delta t}{d} = \sqrt{3} \Delta t^{*} ,
\end{equation} 
so it characterizes the average distance in relation to the particle size.
It is proportional to $\Delta t^{*}$.
Because this reduced distance, and, consequently, the reduced time step should be smaller than $1$, a strict limit is imposed to $\Delta t^{*}$.

In our simulations, we used the value $\Delta t^{*}{=}0.01$.
Larger values caused particle overlaps and instabilities in the simulations.
Smaller values were needed at extreme values of $(\mu^{\mathrm{appl}})^{*}$.

\section{Results and Discussion}
\label{sec:results}

In this study, we report an analysis of the dynamics of chain formation over a wide range of parameter space. 
We study dependence on $N$, $\rho^{*}$, $\gamma^{*}$, $\alpha^{*}$, and $(\mu^{\mathrm{appl}})^{*}{=}\alpha^{*}E_{\mathrm{appl}}^{*}$.
We perform $M_{0}$ time steps in the absence of applied electric field ($E_{\mathrm{appl}}{=}0$), and $M_{\mathrm{E}}$ time steps in the presence of it in order to study the dynamics of chain formation after the electric field is switched on.

We show values of block averages (denoted by $\langle \dots \rangle_{\mathrm{b}}$ for various physical quantities (see the subsequent subsection) as functions of $t^{*}$. 
The length of a block ($M_{\mathrm{b}}$ is the number of time steps in a block), is also a subject of optimization.
If a block is too short, the physical quantities averaged over a block will have bad statistics.
If a block is too long, we lose time resolution and information about the dynamics of the system.

\begin{figure}[t]
	\centering
	\includegraphics[width=0.48\textwidth]{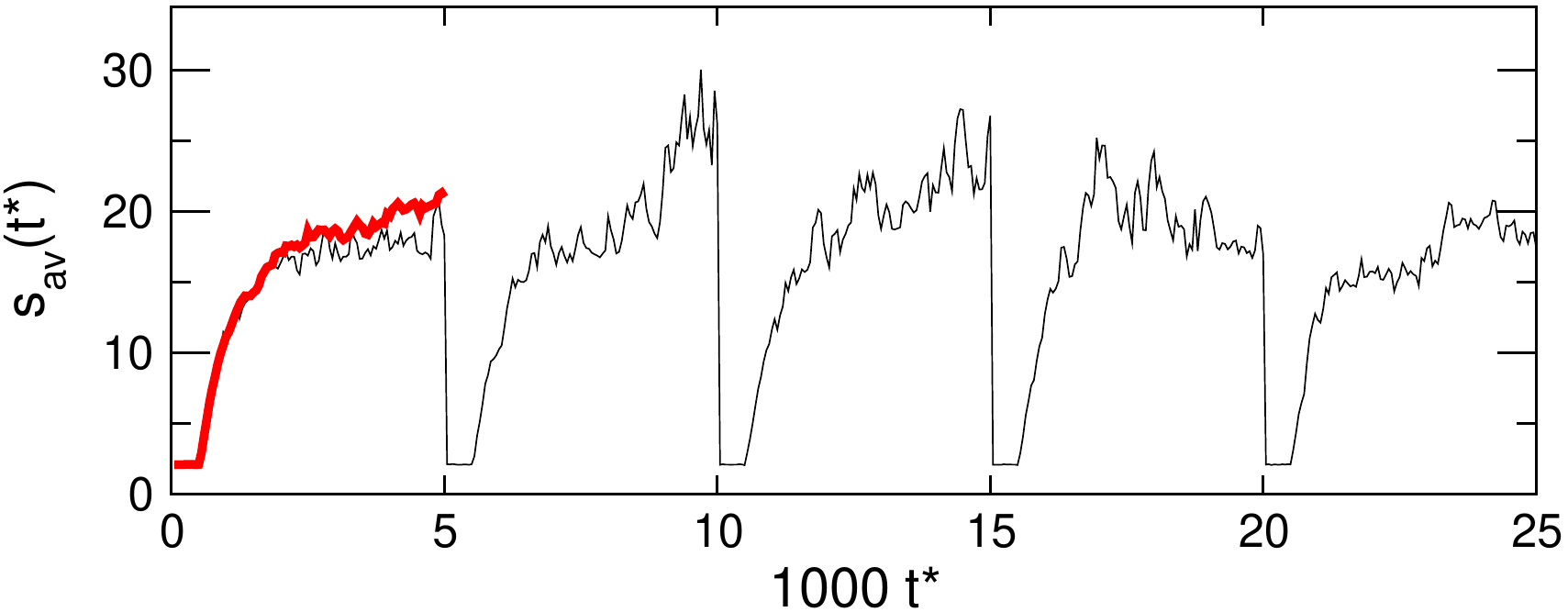}
	\caption{Illustration of the periodic simulation of $M_{0}{=}50,000$ and $M_{E}{=}450,000$ time steps in the absence and in the presence of an applied electric field, respectively ($\Delta t^{*}{=}0.01$). The average chain length is shown as a function of $t^{*}$
	The black line consists of data obtained as averages over blocks of length $M_{\mathrm{b}}{=}5000$ time steps. The think red line shows the average over $20$ periods.}
	\label{fig02}
\end{figure}

Even optimized, the results for the studied quantities are noisy.
Therefore, to improve statistics, we perform several of this $M_{\mathrm{c}}{=}M_{0}{+}M_{\mathrm{E}}$ cycles and average over the cycles (see Fig.~\ref{fig02}).
When we start a cycle over, we restart from a freshly generated initial configuration in a completely disordered state without chains.
This way, the subsequent periods are independent and can be averaged.

In this work, unless we state otherwise, we performed $M_{\mathrm{b}}{=}5000$ time steps in a block, $10$ blocks without field (so, $M_{0}{=}10M_{\mathrm{b}}{=}50,000$), $90$ blocks with field (so $M_{E}{=}90M_{\mathrm{b}}{=}450,000$), and $20$ such periods.
$M_{0}{=}50,000$ and $M_{E}{=}450,000$ time steps corresponds to lengths $t^{*}{=}500$ and $4500$ in reduced time ($\Delta t^{*}{=}0.01$). 
Such a simulation was altogether $10$ million time steps long.

\subsection{Quantities studied}

We can characterize the time dependence of chain formation with several physical quantities. 
Here, we briefly list them, while a detailed description and analysis is found in Ref.~\cite{fertig_hjic_2020}.

In chains, particles are aligned into head-to-tail positions along the $z$-axis that is a lowest-energy configuration.
The one-particle dipolar energies $(u^{\mathrm{appl}})^{*}{=}(U^{\mathrm{appl}})^{*}/N$ and $(u^{\mathrm{dip}})^{*}=(U^{\mathrm{dip}})^{*}/N$ (they are different only if particle-particle polarization is present), therefore, are good indicators of chain formation.

When the particles are ``frozen'' into chains, their mobility characterized by the isotropic diffusion constant decreases that is computed as the slope of the mean square displacement (MSD) as a function of time: 
\begin{equation}
D(t_{\mathrm{b}})= \frac{\langle \mathbf{r}^{2}(t)\rangle_{\mathrm{b}}}{2 \Delta t_{\mathrm{b}}},
\end{equation} 
where $t_{\mathrm{b}}$ is the time at the beginning of a block, and $\Delta t_{\mathrm{b}}{=}M_{\mathrm{b}}\Delta t$ is the length of the block.
(From now on, when we talk about a value of a quantity at a given time $t$, we always mean the average over a block at time $t_{\mathrm{b}}$.) 
Note that $D$ is an approximate value obtained over a block of limited length (see Fig.~6 of Ref.~\cite{fertig_hjic_2020}).
The exact equilibrium diffusion constant would be obtained in the limit of $\Delta t_{\mathrm{b}}{\rightarrow} \infty$.
Even though approximate, $D(t_{\mathrm{b}})$ characterizes chain formation.

The chain formation can be directly followed by identifying chains in every configuration. 
From the number of chains of length $s$, $n_{s}$, for a configuration, we can compute the average chain length as
\begin{equation}
 s_{\mathrm{av}} =\frac{\sum_{s}sn_{s}}{\sum_{s}n_{s}}.
 \label{eq:l}
\end{equation} 
This quantity then can be further averaged over time steps in a block providing time dependence, $s_{\mathrm{av}} (t)$.
We define two particles being in the same chain if they are closer to each other than a predefined distance:  $r_{ij} {<} \lambda_{\mathrm{g}} d $. 
For $\lambda_{\mathrm{g}}$ we use the value $1.2$ in this study, but other numbers produce the same dynamics although with quantitatively different results for $n_{s}$.~\cite{fertig_hjic_2020}
This is a geometrical definition. 
An energetic definition is also possible, but it gives qualitatively the same result as the geometrical one (see Fig.~7 of Ref.~\cite{fertig_hjic_2020}).

The chain distribution $n_{s}(t)$ means a lot of data as a function of $s$ and $t$ that are not easy to visualize. 
Examples are shown in Figs.~8 and 9 of Ref.~\cite{fertig_hjic_2020}.
Here, we will plot $n_{s}(t)$ both as a function of $s$ at fixed $t$ and as a function of $t$ at fixed $s$.

The structure of a fluid can be quantitatively characterized by pair distribution functions.
Although there are various projections of the full pair distribution function in the series expansion of rotational invariants, here we study only the radial distribution function (RDF), $g(r)$, that describes the probability that another particle is found at a distance $r$ from a central particle.
Peaks in $g(r)$ represent probable distances for small $r$, while $g(r){\rightarrow} 1$ when $r{\rightarrow} \infty$ in a fluid phase.
Aggregation of particles increases these peaks.

Although we also compute other distribution functions in the simulations, the conclusions that we can draw from them are not different from those drawn from $g(r)$.

\subsection{Characteristic times}
\label{sec:tau}

Our results will show that the dynamics of the system can be loosely characterized with two processes, a faster one and a slower one.
The faster one can be associated with the formation of chains either from integration from shorter elements or disintegration of longer chains.
The slower one can be associated with the disappearance of chains either from aggregation into even longer chains (or into columnar structures, where chains stick together) or from their disintegration into shorter chains.

In both cases, association and dissociation have the same rate at equilibrium, but at the beginning, it is the association process that dominates. 
The fast subprocess is fast because its rate is determined by the abundance of lone particles (e.g., monomers, $s{=}1$), dimers ($s{=}2$), and short chains. 
The slow subprocess is slow because its rate is determined by the mobility of longer chains or their ability to find monomers or other short chains that they could incorporate. 

The idea to fit bi-exponential functions to the time-dependent function of any physical quantity that we can squeeze out of our simulations naturally arises:
\begin{equation}
 f(t) = a_{0} + a_{1}e^{-t/\tau_{1}} + a_{2}e^{-t/\tau_{2}}.
\end{equation} 
It is clear that using such a function is an approximation. 
The brute force approach would be writing up $N$ coupled differential equations for the components of the system where the components are the chains:
\begin{eqnarray}
 \frac{n_{s}(t)}{dt}  &  = k_{1,s-1}n_{1}n_{s-1} + \dots + k_{s/2,s-s/2}n_{s/2}n_{s-s/2} + \nonumber \\
 & + k'_{s+1,1}n_{s+1} + \dots + k'_{N,N-s}n_{N} - \nonumber \\
 & - (k'_{1,s-1} + \dots + k'_{s/2,s}) n_{s} - \nonumber \\ 
 & - k_{1,s}n_{1}n_{s} -\dots - k_{N-s,s}n_{N-s}n_{s}
  \label{eq:diffeq}
\end{eqnarray}
for every $s{=}1,\dots ,N$.
Here, the first row expresses processes producing chain $s$ by associations from shorter chains $i$ and $j$ with rate constant $k_{i,j}$ with $i{+}j{=}s$. 
The second row expresses processes producing chain $s$ by dissociations from longer chains $i$ with rate constant $k'_{i,i-s}$. 
The third row expresses processes consuming chain $s$ by its dissociation into shorter chains $i$ and $s-i$ with rate constant $k'_{i,s-i}$. 
The fourth row expresses processes consuming chain $s$ by associations with another chain $i$ with rate constant $k_{i,s}$.
Terms containing $s/2$ assume that $s$ is even; for odd $s$ values $(s-1)/2$ should be there.

Solving this system of equations analytically is practically impossible for sensible particle numbers.
Even solving numerically and relating rate constants to data extracted from a simulation is a computational nightmare due to the large number of data that should be visualized and the statistical noise that burden the data.

Therefore, we use the heuristic approach of the bi-exponential fit and characterize the dynamics of the system with the resulting time constants, $\tau_{1}$ and $\tau_{2}$.
The accuracy of the fit is surprisingly good in most cases even though many of the subprocesses in Eq.~\ref{eq:diffeq} are not first-order.

A further advantage of this approach is that we can perform the same kind of fitting for experimental data and relate our model calculations to experimental reality.
For example, the same bi-exponential fit was used by Horv\'{a}th and Szalai~\cite{horvath_pre_2012,horvath_pre_2015} who studied the dielectric response of ER fluids. 
 
\subsection{The importance of particle-particle polarization}

The simulation studies in the ER literature that we aware of have been done for the case when only the dipoles induced by the applied field, $\mu^{\mathrm{appl}}{=}\alpha E_{\mathrm{appl}}$, were used in the calculations, while the dipoles induced by all the other particles, $\mu^{\mathrm{part}}$, were ignored.

\begin{figure*}[t]
	\centering
	\textsf{(A)}\includegraphics[height=0.34\textwidth]{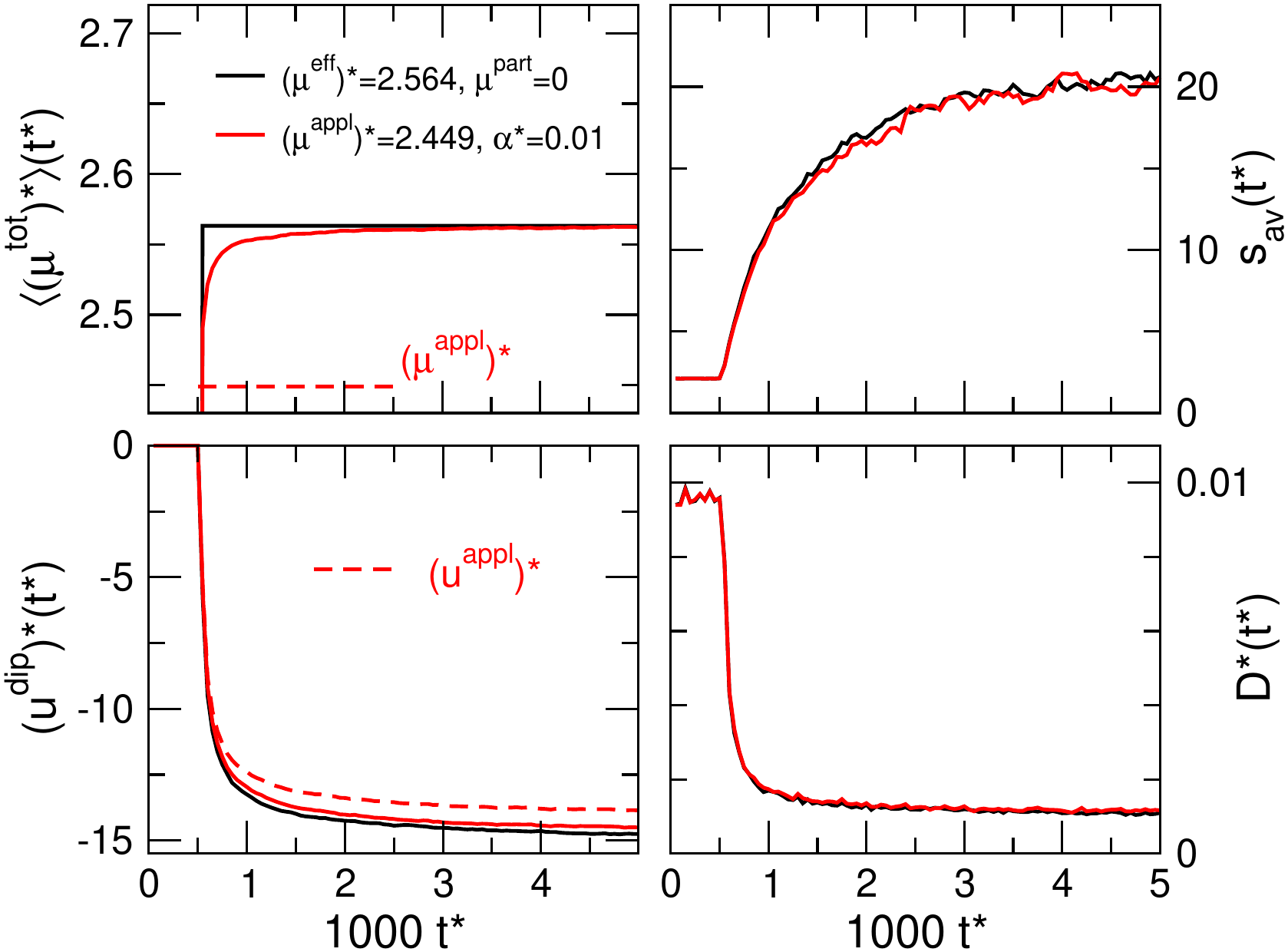}\hspace{0.2cm}
	\textsf{(B)}\includegraphics[height=0.34\textwidth]{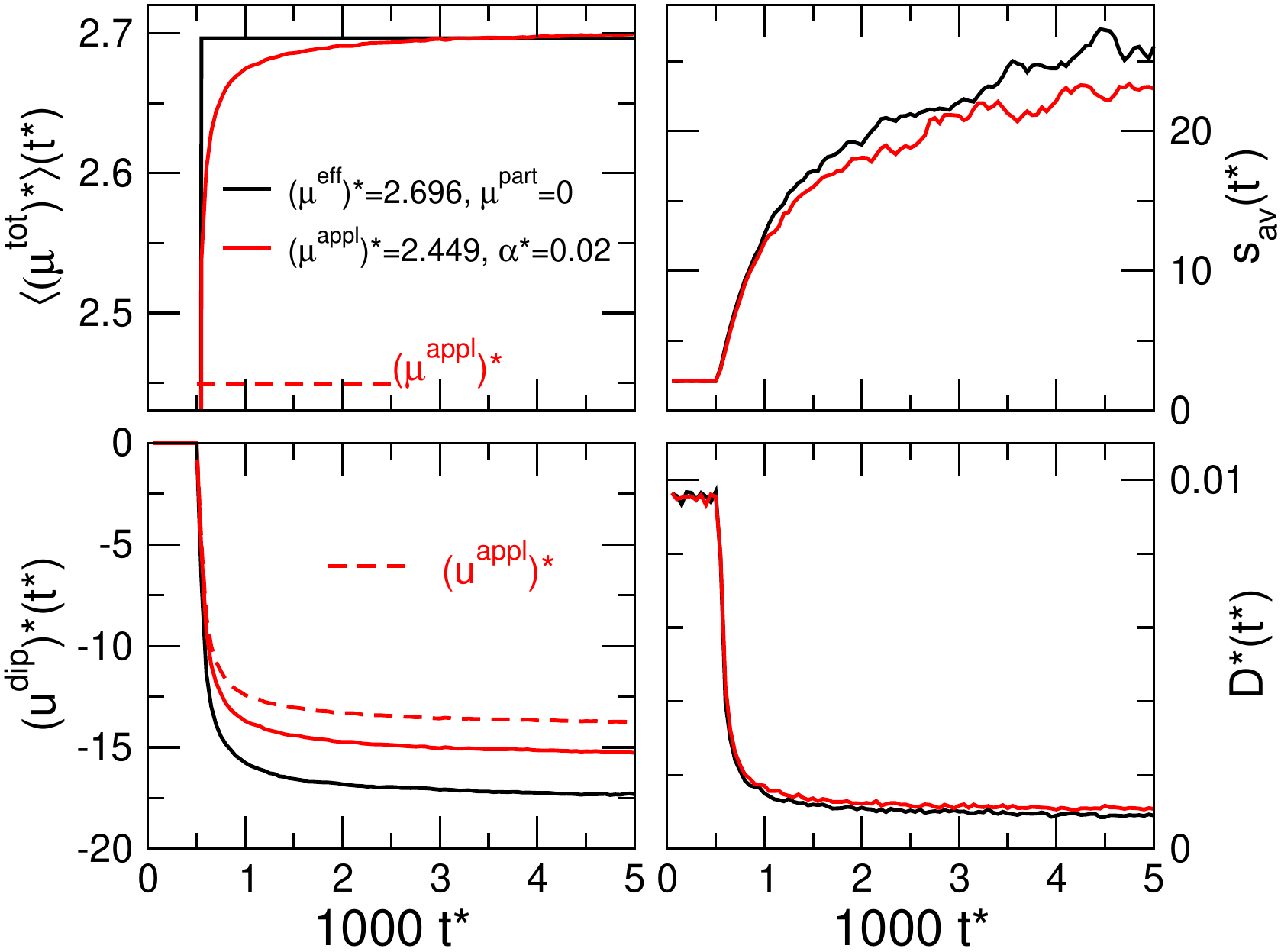}
	\caption{Preliminary results for an ER fluid where polarization of the particles by other particles is taken into account ($\mu^{\mathrm{part}}{\ne} 0$).  Red curves show the results for this case on the examples of various quantities: average chain length, $s_{\mathrm{av}}$, square of the total dipole moment, $\mu{=}\mu^{\mathrm{appl}}{+}\mu^{\mathrm{part}}$, total one-particle dipolar energy, $u^{\mathrm{dip}}{=}u^{\mathrm{appl}}{+}u^{\mathrm{part}}$ (here, $u^{\mathrm{appl}}$ is also shown with dotted line), and the diffusion constant, $D$. These quantities are shown in reduced units. The red curves refer to the case of (A) $\alpha^{*}{=}0.01$  and (B) $\alpha^{*}{=}0.02$ that correspond to $(\mu^{\mathrm{appl}})^{*}{=}\sqrt{6}=2.449$. The values obtained for the total dipole moment are (A) $\langle (\mu^{\mathrm{tot}})^{*}\rangle {=}2.564$ and (B) $\langle (\mu^{\mathrm{tot}})^{*}\rangle {=}2.696$ as ensemble averages. These values were used as effective dipole moments in the simulations without particle-particle polarization (black curves): $(\mu^{\mathrm{eff}})^{*}{=}\langle (\mu^{\mathrm{tot}})^{*}\rangle$. Simulation parameters are $N{=}256$, $\gamma^{*}{=}100$, and $\rho^{*}{=}0.05$. }
	\label{fig03}
\end{figure*}

Here, we show some preliminary results to get an impression of how this approximation works.
Fig.~\ref{fig03} shows the time dependence of various quantities for two pairs of simulations.

In one pair of the simulations, the reduced dipole moment induced by $E_{\mathrm{appl}}$ was kept constant at the value $(\mu^{\mathrm{appl}})^{*}{=}\sqrt{6}{\approx} 2.449$.
Two simulations were performed for $\alpha^{*}{=}0.01$ (Fig.~\ref{fig03}A) and $\alpha^{*}{=}0.02$ (Fig.~\ref{fig03}B). 
The corresponding reduced electric field strengths, $E_{\mathrm{appl}}^{*}$, and the resulting dipole moments $(\mu^{\mathrm{part}})^{*}$ and $(\mu^{\mathrm{tot}})^{*}$ are collected in Table \ref{tab:induction}.
The red curves in Fig.~\ref{fig03} show the results for this case.

In the other pair of the simulations, we used the resulting total dipole moment of the previous simulation as an effective dipole moment, $(\mu^{\mathrm{eff}})^{*}{=}2.563$ and $2.696$, but we ignored the particle-particle polarization ($\mu^{\mathrm{part}}{=}0$ and $U^{\mathrm{part}}{=}0$).
The black curves in Fig.~\ref{fig03} show the results for this case.

\begin{table}[b]
\caption{Simulation parameters for the cases (Figs.~\ref{fig03}A and B) including particle-particle polarization.}
\label{tab:induction}
 	\def\arraystretch{1.3}
 	\centering
\begin{tabular}{l|ll} \hline 
Parameter & Fig.~\ref{fig03}A  & Fig.~\ref{fig03}B \\ \hline
$(\mu^{\mathrm{appl}})^{*}$ & 2.449 & 2.449 \\
$\alpha^{*}$ & 0.01 & 0.02 \\
$E_{\mathrm{appl}}^{*}$ & 244.9 & 122.45 \\
$\langle (\mu^{\mathrm{part}})^{*}\rangle$ & 0.114 & 0.247 \\ 
$\langle (\mu^{\mathrm{tot}})^{*}\rangle$ & 2.563 & 2.696 \\ \hline 
\end{tabular}
\end{table}

The top-left panels of Figs.~\ref{fig03}A and B illustrate the relation of the two kinds of simulations.
While the total dipole moment jumps to the $(\mu^{\mathrm{eff}})^{*}$ value abruptly as the applied field is switched on in the $\mu^{\mathrm{part}}{=}0$ case (black curves), it gradually approaches the same value in the case when particles polarize each other and it takes some time for them to aggregate and polarize each other (red lines).
The total dipole moment is also a key quantity for the calculation of the dielectric constant 
\begin{equation}
 \frac{\epsilon-1}{\epsilon+2} = \frac{4\pi}{3} \alpha \rho \frac{\langle \mu^{\mathrm{tot}}\rangle }{\mu^{\mathrm{appl}}} = \frac{4\pi}{3} \alpha \rho (1+S),
\end{equation} 
where the correction factor, $S{=}\langle \mu^{\mathrm{part}}\rangle / \mu^{\mathrm{appl}}$, characterizes the deviation from the Clausius-Mosotti equation.~\cite{valisko_jcp_2009}
The dielectric constant is a well-measurable quantity,~\cite{horvath_pre_2015,horvath_pre_2012} and, therefore, is of crucial importance. 
We will devote a separate paper to its investigation.

The other quantity, where we observe deviation between the two kinds of simulations is the energy (bottom-left panels).
This deviation follows from the way $U^{\mathrm{appl}}$ and $U^{\mathrm{part}}$ are defined; dipole moments $\mu_{i}^{\mathrm{appl}}$ and $\mu^{\mathrm{part}}_{i}$ are separate dipole moments. 
There are terms that seem to be missing from Eqs.~\ref{eq:uappl} and \ref{eq:upart}.
The missing terms result from collecting all the $\mu^{\mathrm{appl}}-\mu^{\mathrm{appl}}$, $\mu^{\mathrm{appl}}-\mu^{\mathrm{part}}$, and $\mu^{\mathrm{part}}-\mu^{\mathrm{part}}$ interactions and deducting the self-polarization term.~\cite{predota_mp_2002}
The black curve can be recovered with proper rescaling.

The diffusion constant (bottom-right panels) behaves the same way in the two kinds of simulations.
This result implies that the particles influence each other's mobility when they are already close to each other so the induced dipoles, $\mu^{\mathrm{part}}$, are formed.

The average chain length, $s_{\mathrm{av}}$, is shown in the top-right panels.
Its behavior is practically the same for the two kinds of simulation in the case of $\alpha^{*}{=}0.01$, while deviations occur for $\alpha^{*}{=}0.02$.
Larger $\alpha^{*}$ values produce even larger deviations (data not shown).

From these preliminary results we can conclude that explicit consideration of particle-particle polarization becomes important as $\alpha^{*}$ increases, and, probably, as $\rho^{*}$ increases.
We believe that this is an important result especially in the light of the fact that we have not found any single simulation study in the literature where particle-particle polarization was taken into account.
We will devote a separate paper to this effect, while in this paper we stay with the ``traditional'' approach, namely, we use the approximations $\mu^{\mathrm{part}}_{i}=0$, $U^{\mathrm{part}}=0$, and $\mathbf{f}_{ij}^{\mathrm{part}}=0$.

In this approximation, we cannot have independent values for $\alpha^{*}$ and $E_{\mathrm{appl}}^{*}$ separately, only for their product: $(\mu^{\mathrm{appl}})^{*}{=}\alpha^{*}E_{\mathrm{appl}}^{*}$.
Therefore, from now on, our independent variable will be $(\mu^{\mathrm{appl}})^{*}$ (or its square, see Eq.~\ref{eq:squaredip}), so, to simplify notation, we will denote $\mu^{\mathrm{appl}}$ simply with $\mu$.
One should, however, keep in mind that this \textit{dipole moment is the result of polarization by an applied field}.

\subsection{The effect of the number of particles}
\label{sec:Ndep}

\begin{figure}[t]
	\centering
	\includegraphics[width=0.48\textwidth]{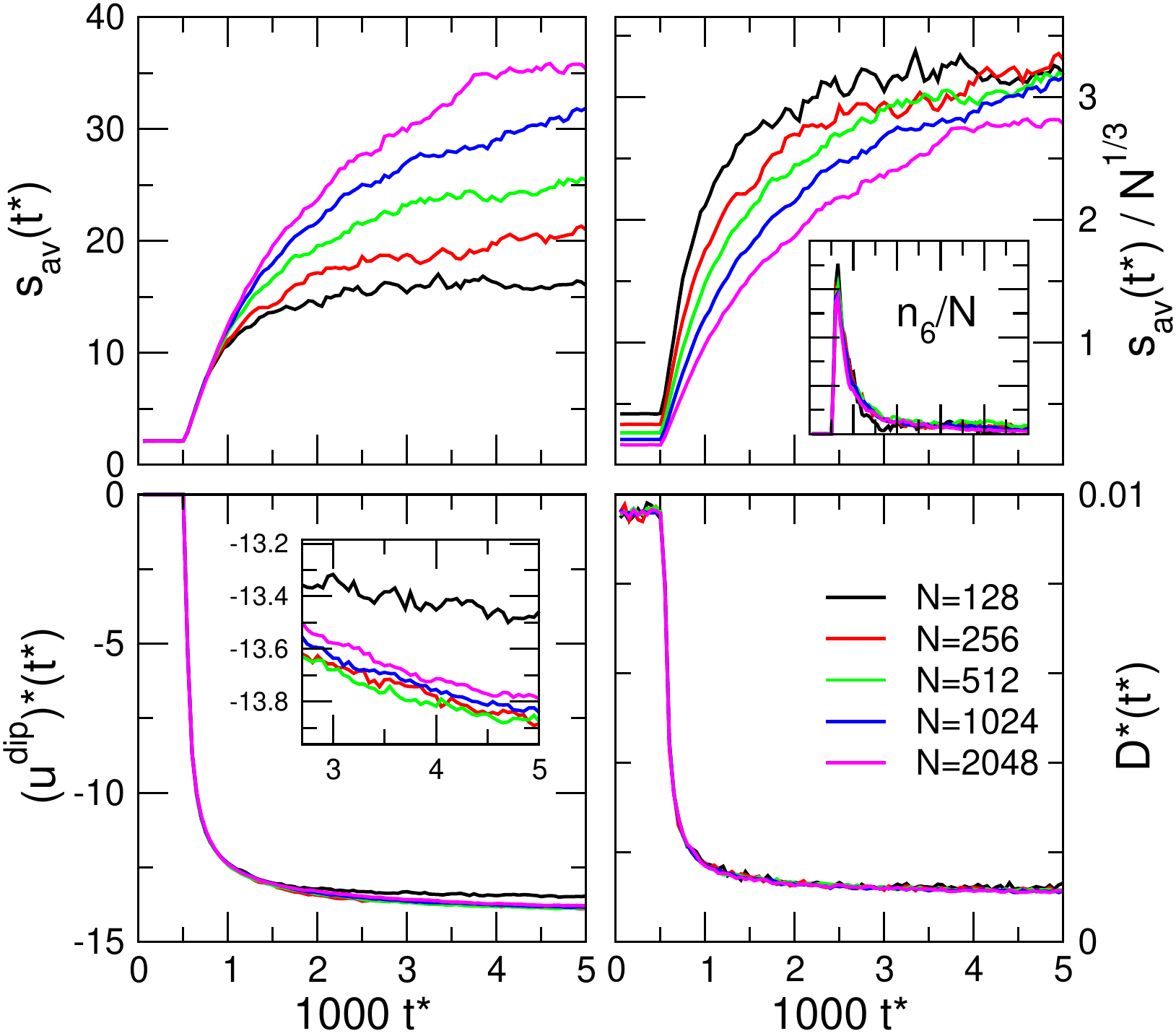}
	\caption{System size analysis. Average chain length (top-left panel), average chain length normalized by the cube root of particle number (top-right panel), one-particle dipolar energy (bottom-left panel), and diffusion constant (bottom-right panel) as functions of $t^{*}$ for different particle numbers, $N$. In the inset of the top-right panel, the time evolution of the number of chains with length $6$ normalized by the particle number are shown. Simulation parameters are $(\mu^{*})^{2}{=}6$, $\gamma^{*}{=}100$, and $\rho^{*}{=}0.05$.}
	\label{fig04}
\end{figure}

Most of the simulations in this paper will be shown for $N{=}256$, so we need to justify why this value is appropriate to characterize larger (more realistic) system sizes.
Fig.~\ref{fig04} shows the time dependence of the quantities already discussed at Fig.~\ref{fig03} for different system sizes for a dipole moment where considerable chain formation is observed, $(\mu^{*})^{2}{=}6$.

The time dependence of the one-particle dipolar energy and the diffusion constant practically do not depend on the number of particles if $N{\geq}256$ (bottom panels).
The inset of the bottom-left panel shows that $N{=}128$ is probably too small, but the behavior is still the same qualitatively.

The behavior of the average chain length, on the other hand, does depend on $N$ (top-left panel). 
The left hand side panel in the top row shows that the equilibrium limit of $s_{\mathrm{av}}$ depends on the size of the simulation cell: larger cells can enclose longer chains.
The analyses of chains of various lengths is necessarily system size dependent.
Because the length of the chain overarching the cell scales with $L{\sim}N^{1/3}$ and the number of short chains scales with $N$, it is expected that we can extrapolate our results for a necessarily small system size to larger ones.

If we scale the average chain length with $N^{1/3}$, we obtain that the equilibrium limit scales with $N^{1/3}$ (top-right panel). 
This result harmonizes with the result (see later) that the average chain length  is dominated by the length of the chain that overarches the cell for $(\mu^{*})^{2}{=}6$; there is a peak in the chain-length distribution at that value.
This value depends on the width of the simulation cell, $L$.

The inset shows that the number of short chains scales with $N$. 
This is also logical.
The dynamics of shorter and longer chains, therefore, has different $N$ dependence, but it is possible to draw conclusions for larger systems from our small-system simulations.

System size in a computer simulation is necessarily finite. 
We generally need to compromise between large computation time and system-size artifacts.
Here, we use $N{=}256$ for the rest of the paper.

\subsection{The effect of friction coefficient}
\label{eq:gamma}

\begin{figure}[t]
	\centering
	\includegraphics[width=0.47\textwidth]{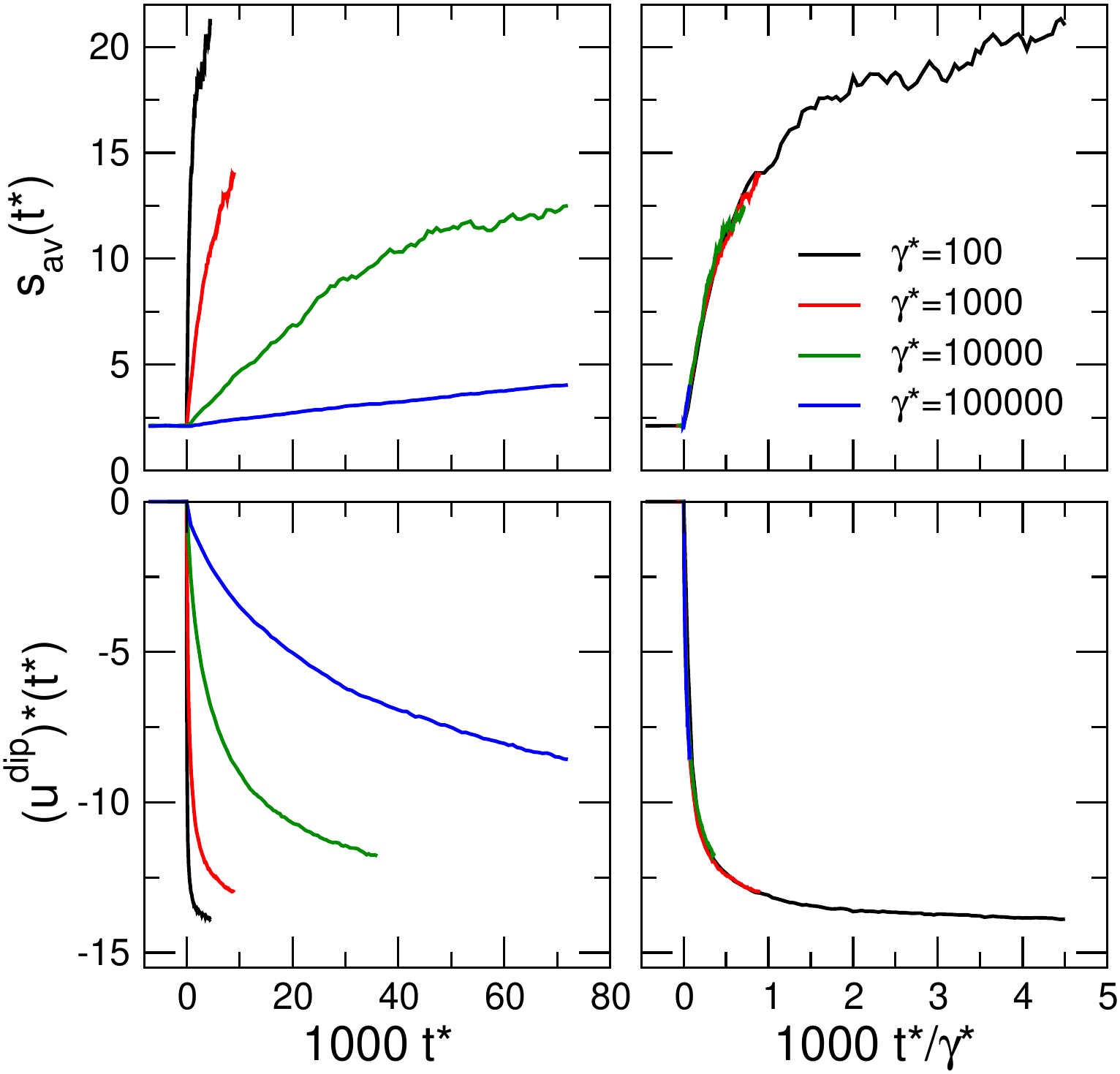}
	\caption{The average chain length (top row) and one-particle reduced dipolar energy (bottom row) as functions of $t^{*}$ (left column) and $t^{*}/\gamma^{*}$ (right column) for various values of the reduced friction coefficient, $\gamma^{*}$, are shown. Simulation parameters are $N{=}256$, $(\mu^{*})^{2}{=}6$, and $\rho^{*}{=}0.05$. The length of a block is $N_{\mathrm{b}}{=}5000$ for $\gamma^{*}{=}{100}$, $N_{\mathrm{b}}{=}10,000$ for $\gamma^{*}{=}{1000}$, and $N_{\mathrm{b}}{=}80,000$ for $\gamma^{*}{=}{10,000}$ and $\gamma^{*}{=}100,000$.}
	\label{fig05}
\end{figure}

The reduced friction coefficient tunes the rate at which the particles diffuse (Eq.~\ref{eq:einsten}).
Larger $\gamma^{*}$ results in a slower evolving simulation as shown by the left panels of Fig.~\ref{fig05}.
If we scale the reduced time with $\gamma^{*}$ as shown in the right panels of Fig.~\ref{fig05}, the curves for the average chain length and the dipolar energy coincide.
This is not true, however, for the diffusion constant, see Fig.~13 of Ref.~\cite{fertig_hjic_2020}.

\begin{figure}[t]
	\centering
	\includegraphics[width=0.35\textwidth]{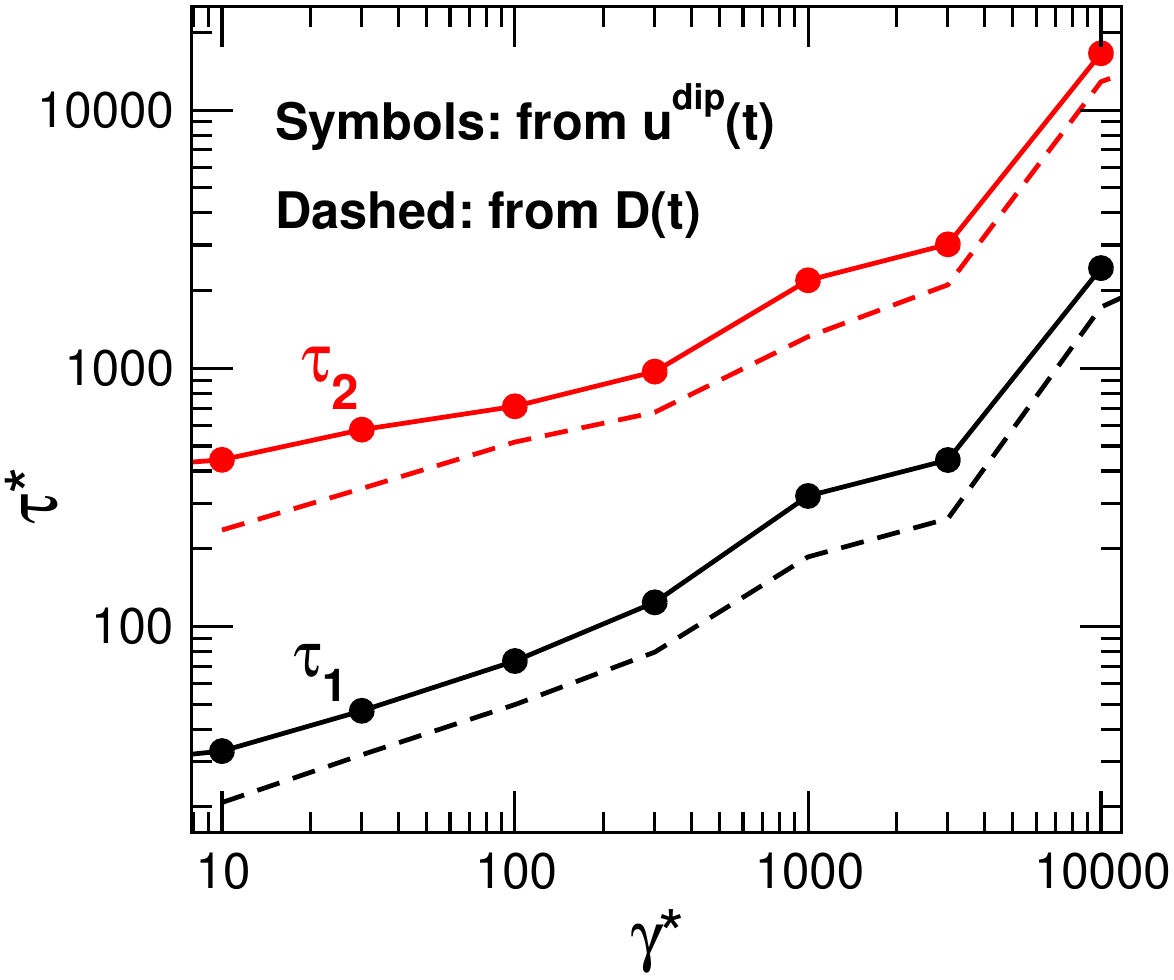}
	\caption{Time constants obtained from bi-exponential fits to $(u^{\mathrm{dip}})^{*}(t^{*})$ (symbols) and $D^{*}(t^{*})$ (dashed lines) as functions of the friction coefficient, $\gamma^{*}$. Parameters are the same as in Fig.~\ref{fig05}.}
	\label{fig06}
\end{figure}

This result implies that we can extrapolate to large values of $\gamma^{*}$ from simulations performed for a small value of $\gamma^{*}$.
This statement is also supported by Fig.~\ref{fig06} that shows the characteristic times as functions of $\gamma^{*}$.
They have a monotonic and smooth $\gamma^{*}$ dependence.
This figure indicates that there is an order of magnitude difference between $\tau_{2}^{*}$ and $\tau_{1}^{*}$.

\subsection{The effect of dipole moment (electric field)}
\label{sec:dipolemoment}

The most important issue is the dependence of chain formation on the applied electric field because $E_{\mathrm{appl}}$ is the major external control parameter.
$E_{\mathrm{appl}}$-dependence means $\mu$-dependence in the absence of particle-particle polarization, so we will use that nomenclature from now on and talk about $\mu^{*}$-dependence.
Also, $\mu^{*}$ (and, especially, $(\mu^{*})^{2}$, see Eq.~\ref{eq:squaredip}) expresses the strength of the ordering effect of the applied field in relation to the disordering effect of thermal motion.

\begin{figure}[t]
	\centering
	\includegraphics[width=0.48\textwidth]{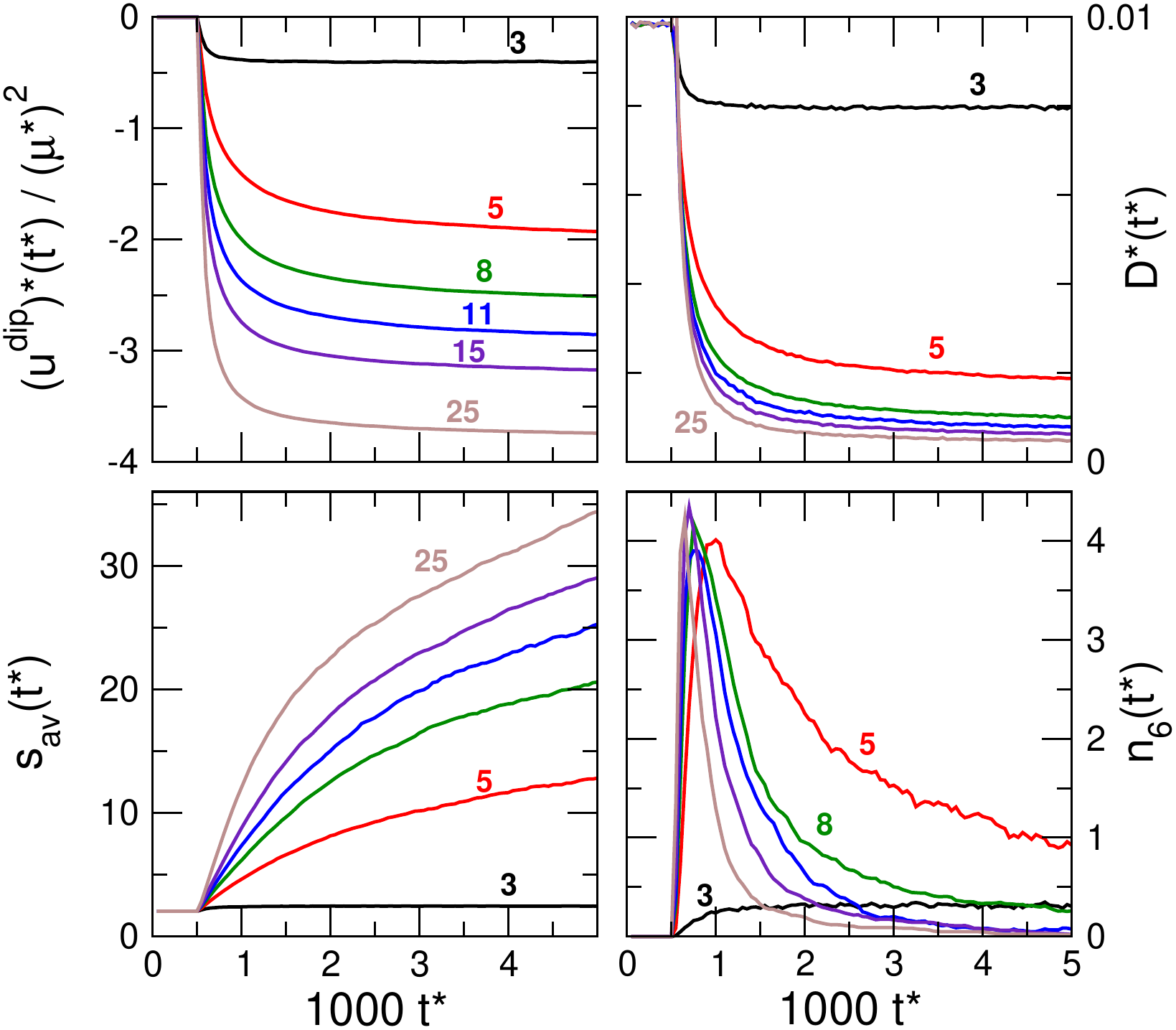}
	\caption{One particle dipolar energy (top-left panel), diffusion constant (top-right panel), average chain length (bottom-left panel), and the number of chains of length $6$ (bottom-right panel) as functions of $t^{*}$ for different values of $(\mu^{*})^{2}$ (3,5,8,11,15,25). Simulation parameters are $N{=}256$, $\gamma^{*}{=}100$, and $\rho^{*}{=}0.02$. In these simulations, we peformed $200$ cycles of the $M_{0}{+}M_{E}$ period. }
	\label{fig07}
\end{figure}

We performed simulations for $(\mu^{*})^{2}{=}3$, $5$, $8$, $11$, $15$, and $25$.
Fig.~\ref{fig07} shows the time dependence of the dipolar energy, the diffusion constant, the average chain length, and the number of chains of length $s{=}6$.
The energy decreases to deeper values as $(\mu^{*})^{2}$ increases, because the dipolar energy is proportional to $\mu^{2}$ (see Eq.~\ref{eq:udippair}).
The dipolar energy, however, decreases disproportionately with $\mu^{2}$ that is well visible in Fig.~\ref{fig07}, where we plot the dipolar energy normalized by $(\mu^{*})^{2}$.

\begin{figure*}[t]
	\centering
 	\includegraphics[width=0.99\textwidth]{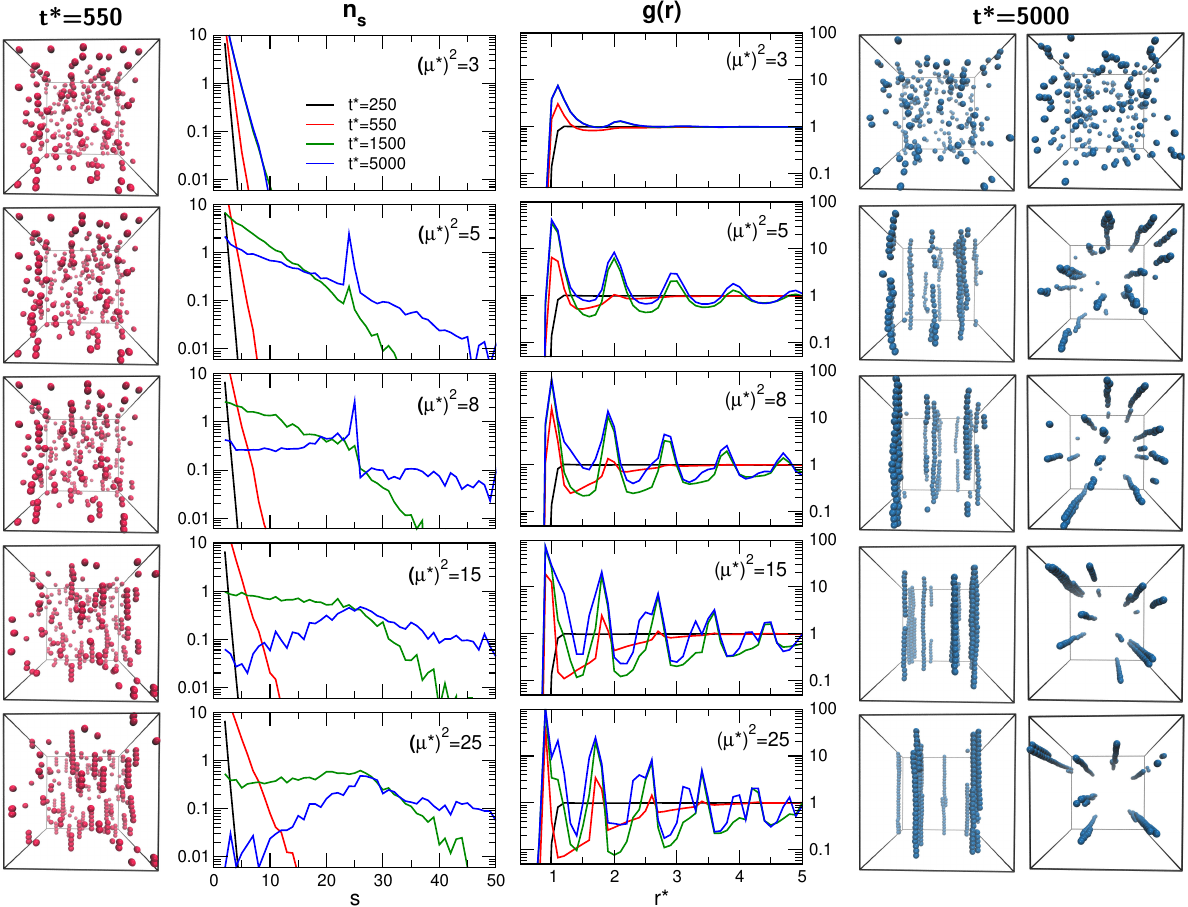}
	\caption{Chain length distributions (second column) and radial distribution functions (third column) for different values of $(\mu^{*})^{2}$ (increasing from $3$ to $25$ from top to bottom). The black curves refer to a block in the absence of $E_{\mathrm{appl}}$ ($t^{*}{=}250$). The red curves refer to a block at the beginning of the period in the presence of $E_{\mathrm{appl}}$ when chains start forming ($t^{*}{=}550$). The green curves refer to a block during chain formation ($t^{*}{=}1500$). The blue curves refer to a block at the end of the period in the presence of $E_{\mathrm{appl}}$ when chains (and possibly aggregation of chains) have formed ($t^{*}{=}5000$). Additionally, for each $(\mu^{*})^2$, snapshots from the simulations are shown at $t^{*}{=}550$ (first column, front view of the simulation cell) and $t^{*}{=}5000$ (fourth and fifth columns, front and top views of the simulation cell, respectively). Parameters are the same as in Fig.~\ref{fig07}.}
	\label{fig08}
\end{figure*}

This disproportionate decrease of the dipolar energy with $(\mu^{*})^{2}$ is explained by the increased aggregation of the particles caused by the stronger interactions. 
The dipolar interaction can be so strong (reduced dipole moments $(\mu^{*})^{2}{=}11$ and above are really large) that the attraction in the head-to-tail position overcomes the repulsion of the core potential.
As a result, the particles can get closer to each other than $r{=}d$.
This is directly shown by the first peaks of the RDF shifting towards smaller $r^{*}$ values as $(\mu^{*})^{2}$ increases (Fig.~\ref{fig08}).

Increased order is also shown by the increasing average chain length (bottom-left panel of Fig.~\ref{fig07}), the decreasing diffusion coefficient (top-right panel of Fig.~\ref{fig07}), the increasing peaks of the RDFs (Fig.~\ref{fig08}), and the snapshots in Fig.~\ref{fig08}. 
The formation of larger aggregates and the disappearance of short chains is shown by the behavior of $n_{6}(t^{*})$ that declines to zero for $(\mu^{*})^{2}$ values larger than $8$ (bottom-right panel of Fig.~\ref{fig07}).

\begin{figure*}[t]
	\centering
	\includegraphics[width=0.99\textwidth]{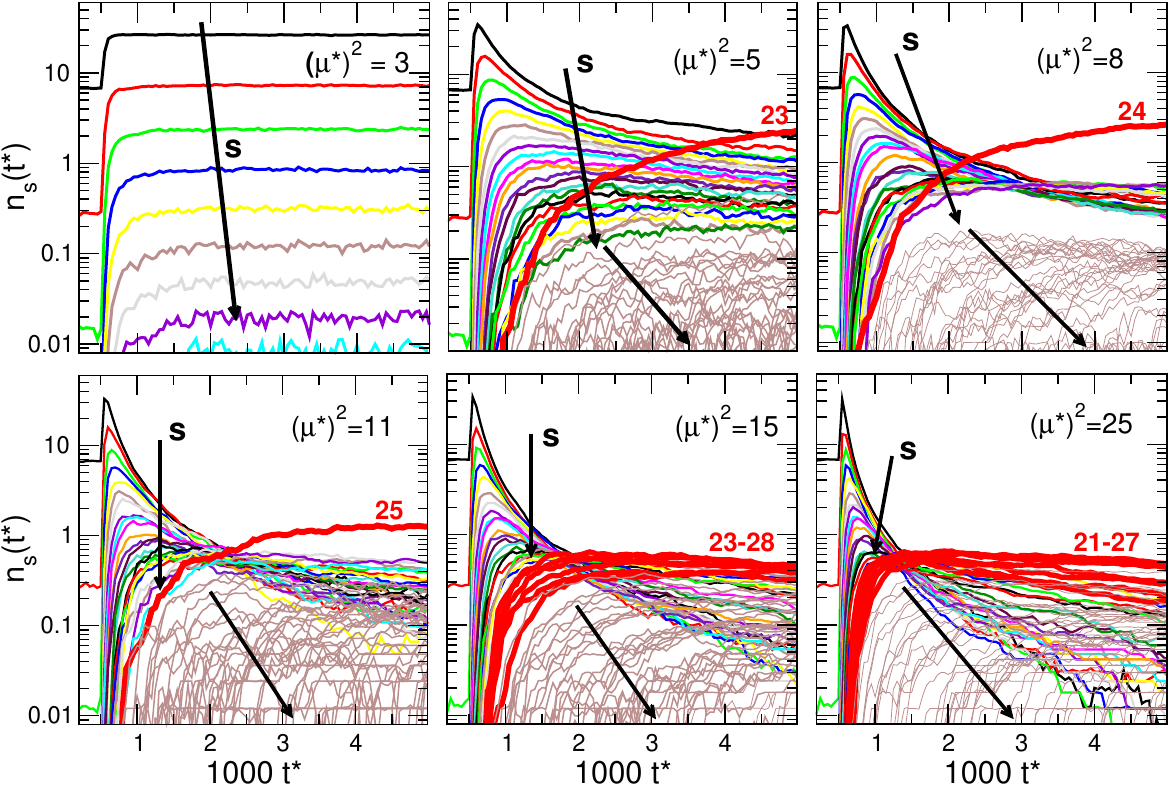}
	\caption{The number of chains of specific lengths ($s{\geq} 2$) averaged over blocks as a function of time, $t^{*}$. The panels refer to dipole moments $(\mu^{*})^{2}{=}3$, $5$, $8$, $11$, $15$ and $25$. Chain length $s$ increases along the arrows. The thick red lines refer to a chain that completely crosses the simulation cell in the $z$ direction. In these simulations ($N{=}256$, $\rho^{*}{=}0.02$), the length of this overarching chain is around the value $n_{L}{=}23$ (for comparison, the width of the simulation cell is $L{=}23.4d$). The color lines refer to the chains of lengths up to $n_{s_{0}}$, while brown lines refer to the chains of lengths above $n_{s_{0}}$. Parameters are the same as in Fig.~\ref{fig07}.}
	\label{fig09}
\end{figure*}

Eventually, chains aggregate into columnar structures when the ordering effect of the applied field is large enough.
The structures formed at large couplings between dipoles, however, get quite close to freezing. 
This is indicated by the quickly declining diffusion constant in Fig.~\ref{fig07}.
Its value, however, never declines to zero which indicates that the system is not frozen.
It rather behaves as a two-dimensional fluid of chains.
This means that the chains, once formed, are quite stable and they diffuse in the $(x,y)$ plane as autonomous entities.
These chains are heavier and less mobile than individual ER particles. 
Also, their kinetic energy is partially stored in a rotation around their $z$-axes. 

The behavior of the chains of various lengths is shown by Fig.~\ref{fig09} that shows the $n_{s}(t^{*})$ vs.\ $t^{*}$ functions for different values of $(\mu^{*})^{2}$ in the different panels.
The $n_{s}(t^{*})$ curves for chains shorter than $s_{0}$ (the length of the chain overarching the cell) are plotted by thin lines of various colors.
The $n_{s_{0}}(t^{*})$ curves are plotted with thick red lines ($(\mu^{*})^{2}{=}5$, $8$, and $11$). 
They are not only one lines, but more for larger values of $(\mu^{*})^{2}$ ($15$ and $25$).
The $n_{s}(t^{*})$ curves for aggregated chains ($s{>}s_{0}$) are plotted with thin brown lines. 
They are very noisy, so it is their common behavior that is meaningful instead of individual ones.

In the case of $(\mu^{*})^{2}{=}3$ we do not observe chain formation.
Short chains may form, but their numbers get smaller as their lengths get larger.
This is also shown by Fig.~\ref{fig08} plotting chain length distributions ($n_{s}$ vs.\ $s$ functions) and radial distribution functions, $g(r^{*})$, for four time moments (actually, time blocks): 
\begin{itemize}
 \item $t^{*}{=}250$, in the absence of $E_{\mathrm{appl}}$ (the WCA fluid, black lines)
 \item $t^{*}{=}550$, at the beginning of the time period in the presence of $E_{\mathrm{appl}}$ when chains just started to form (red lines)
 \item $t^{*}{=}1500$, in the middle of the time period in the presence of $E_{\mathrm{appl}}$ when individual chains are mostly formed (green lines)
\item $t^{*}{=}5000$, at the end of the simulated time period in the presence of $E_{\mathrm{appl}}$ when chains have formed and (for large $\mu^{*}$ values) aggregated (blue lines)
\end{itemize}
For $(\mu^{*})^{2}{=}3$, the $n_{s}$ functions show a simple decreasing behavior, but shifted towards larger $s$ values as time goes by (Fig.~\ref{fig08}).
The $g(r)$ functions show a minimal structure in the presence of $E_{\mathrm{appl}}$ with just two peaks compared to the black line (WCA fluid) that just shows a gas-like behavior.

In the case of $(\mu^{*})^{2}{=}5$, two chains of length $s_{0}$ are present in the simulation cell (on average) as shown by the thick red line.
This curve ``jumps out'' of the crowd of curves of other chains indicating that this chain has a special status among all the chains (higher probability).
Whether this is a simulation artifact can be the subject of further investigation.
If it is an artifact, it is due to the periodic boundary conditions that stabilize this chain because it is practically a loop in which every particle has the lowest energy due to its periodic neighbors.
The fact that this curve ``jumps out'' even for larger system sizes ($N{=}2048$) implies that this might be a real physical effect characteristic even to real system sizes. 
After all, even in a real macroscopic experimental cell, the chains that overarch the slit between the two electrodes can be more stable because they are bound at the two ends.

For $(\mu^{*})^{2}{=}5$, we observe that shorter chains behave as intermediates (see thin colored lines).
First, they are produced from shorter chains, then their number decreases approaching their equilibrium values that decrease with increasing $s$ (Fig.~\ref{fig09}).
The $n_{s}$ distribution now is also decreasing, but now larger $s$ values are possible (Fig.~\ref{fig08}).
The interesting thing is the peak at $s_{0}{=}23$ (beware the logarithmic scale) that indicates the special role of this overarching chain.
The $g(r)$ function clearly indicates strong structuring.
This is an obvious sign of aggregation because the fluid is otherwise low density ($\rho^{*}{=}0.02$).

If we increase the dipole moment to  $(\mu^{*})^{2}{=}8$, we can see similar phenomena except that we observe an interesting gap between the lines below $s_{0}{=}24$ (thin lines with colors) and above it (thin brown lines).
There is a jump corresponding to this gap in the $n_{s}$ vs.\ $s$ function in Fig.~\ref{fig08}.
If there is a phase transition with increasing $\mu$ (we are unsure), it must be somewhere here.
We must be careful with such statements, however.
We need to conduct a more detailed and less noisy investigation for larger systems if we want to be sure about the existence of a phase transition.

\begin{table*}[t]
\caption{Properties of various ER fluids and the corresponding reduced quantities.}
\label{tab:example}
 	\def\arraystretch{1.2}
 	\centering
\begin{tabular}{l|ll|llll|lll} \hline 
Ref. & ER particle & solvent & $d$ & $\epsilon_{\mathrm{in}}$ & $\epsilon_{\mathrm{out}}$ & $\epsilon_{\mathrm{in}}/\epsilon_{\mathrm{out}}$ & $\alpha^{*}$ & $E_{\mathrm{appl}}^{*}$ & $\mu^{*}$  \\ \hline
\cite{whittle_jnnfm_1990} & \makecell{lithium-\\polymethacrylate} & light oil & $1.5{\times} 10^{-5}$ & $30$ & $3$ & $10$ & $0.094$ & $9553$ & $896$ \\
\cite{tao_prl_1994,tao_ijmpb_1994} & alumina & petroleum oil & $10^{-5}$ & $8$ & $2$ & $4$ & $0.063$ & $5200$ & $325$ \\
\cite{klingenberg_langmuir_1990} & poly(methacrylate) & \makecell{chlorinated\\ hydrocarbon} & $10^{-5}$ & $23.36$ & $7.3$ & $3.2$ & $0.053$ & $5200$ & $275$ \\
\cite{cao_jpcb_2006} &  SrCO$_{3}$ & silicon oil & $10^{-6}$         & NA & 2 & $\sim 800$ & $0.125$ & $164.4$ & $20.5$ \\
\cite{cao_jpcb_2006} & SrCO$_{3}$ & silicon oil & $5{\times} 10^{-7}$ & NA & 2 & $\sim 300$ & $0.124$ & $58.1$ & $7.2$ \\
\cite{horvath_pre_2012,horvath_pre_2015} & silica (SiO$_{2}$) & silicon oil & $2{\times} 10^{-8}$ & $4$ & $2.7$ & $1.48$ & $0.017$ & $0.465$ & $0.00804$ \\ \hline
\end{tabular}
\end{table*}

Increasing the dipole moment even further ($(\mu^{*})^{2}{=}11$, $15$, $25$), the gap vanishes.
The thick red line still  ``jumps out'' for $(\mu^{*})^{2}{=}11$. 
For even larger values of $(\mu^{*})^{2}$ ($15$ and $25$), we can find not only a single thick red line, but a collection of thick red lines.
This and the fact that the value of $s_{0}$ increases ($23$, $24$, and so on) as $(\mu^{*})^{2}$ increases can be explained by the fact that the ER particles can get closer to each other when the strong dipolar interaction attracts them.
This is also shown by the $g(r)$ functions in Fig.~\ref{fig08}: the first peak is getting below the $r^{*}{=}1$ contact position of the WCA fluid.
The soft WCA potential is pretty ``hard'' normally, but if there is a strong attractive force balancing it, it becomes ``softer''.
This flexibility allows overarching chains of different lengths.

This phenomenon results in a smaller effective diameter of the particles compared to the value of $d$ used in the WCA potential and with which we define the reduced quantities.
This might be worth taking into account during analyzing our results.

The number of chains longer than $s_{0}$ exceeds the number of chains shorter than $s_{0}$.
This is shown by the brown lines overstepping the colorful lines in Fig.~\ref{fig09} and by the relation of the green and blue curves in Fig.~\ref{fig08}.
Aggregation of chains that corresponds to chains with lengths larger than $s_{0}$ is shown by the ``shoulders'' near the peaks of the $g(r)$ functions in Fig.~\ref{fig08}.

These simulations have been done for a low density ($\rho^{*}{=}0.02$) in order to identify and visualize the chains better.
We have not performed simulations for larger $(\mu^{*})^{2}$ values in this work for the following reasons.
\begin{itemize}
 \item The simulations were problematic computationally; overlap of particles caused these particles to ``shoot apart'' due to the large repulsion. 
 This problem can be solved with smaller $\Delta t^{*}$ or larger $\gamma^{*}$ values that correspond to larger computational time.
 \item The process of chain formation, that is our main interest here, was so abrupt for $\gamma^{*}=100$ used in these simulations that we could not really follow the dynamics. This problem can be solved with larger $\gamma^{*}$ values that, again, corresponds to larger computational time.
\end{itemize}
Therefore, we did not pursue these state points here, instead, we refer them to future studies.

These state points can be especially important if we want to relate our reduced units to real ER fluids. 
Table \ref{tab:example} shows a few examples that we found in the literature.
As far as the value of the reduced dipole moment is concerned, we encountered ER fluids that correspond to much higher and much lower $\mu^{*}$ values than those we work with in our study (for $E_{\mathrm{appl}}{=}10^{6}$ V/m).
For example, the value $\mu^{*}{=}0.00804$,~\cite{horvath_pre_2012,horvath_pre_2015} according to our simulations, is too small to produce chain formation (this small value is due to the nano-size diameter of the particles). 
This is the result of both the low $\epsilon_{\mathrm{in}}/\epsilon_{\mathrm{out}}$ ratio and the small $d$.
The values of $\mu^{*}$ above $20$,~\cite{whittle_jnnfm_1990,tao_prl_1994,tao_ijmpb_1994,klingenberg_langmuir_1990} on the other hand, are so large that chains aggregate into stable structures and the ER particles solidify. 
These large values of $\mu^{*}$ are generally the results of the large particle diameters.

This might be the real experimental situation, but it is hard to study with our simulation methodology. 
Beyond a certain  point, we simulate the movement of chains rather than the movement of particles accumulating to form chains.
Their motion is much slower (see the diffusion constants in Fig.~\ref{fig07}), but it is doable because it is a two-dimensional fluid of chains.
When the chains aggregate, however, we have a solid-like structure that requires special sampling techniques.

The realm of state points (in reduced units) that we simulate in this work, therefore, corresponds to moderate applied fields ($E_{\mathrm{appl}}{\approx} 10^{4}-10^{5}$ V/m) or not too large particles.

\begin{figure}[t]
	\centering
	\includegraphics[width=0.32\textwidth]{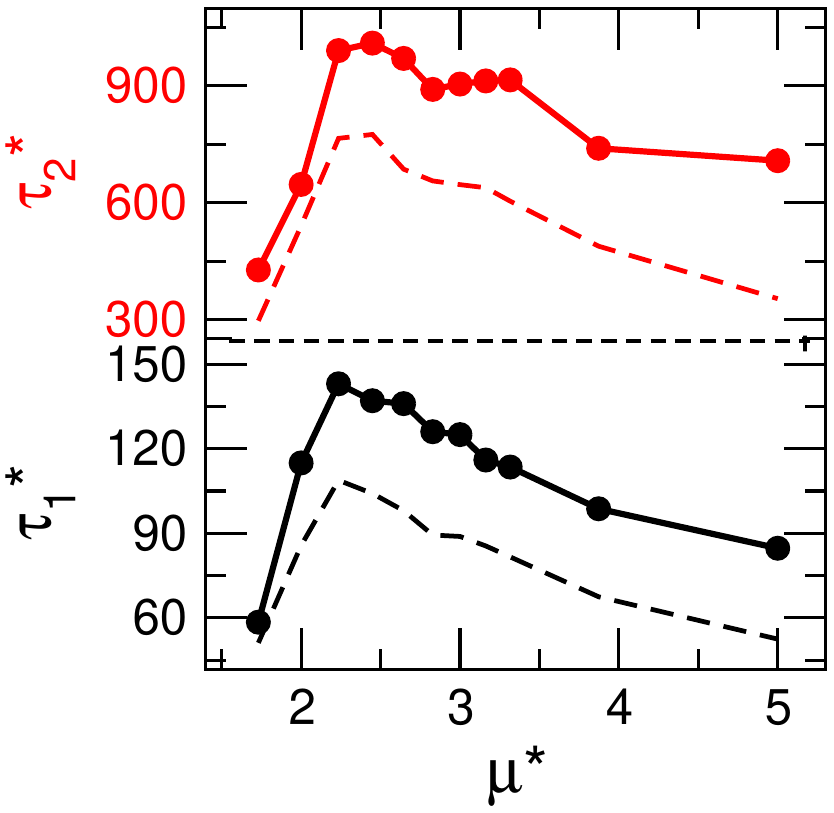}
	\caption{The $\tau^{*}_1$ (black, bottom panel) and $\tau^{*}_2$ (red, top panel) time constants obtained from bi-exponential fits to $(u^{\mathrm{dip}})^{*}(t^{*})$ (symbols) and $D^{*}(t^{*})$ (dashed lines) as functions of $\mu^{*}$. Parameters are the same as in Fig.~\ref{fig07}.}
	\label{fig10}
\end{figure}

Although a clear-cut proof for phase transition has not been found, we observe an interesting maximum in the characteristic times, $\tau_{1}^{*}$ and $\tau_{2}^{*}$, as functions of $\mu^{*}$ in Fig.~\ref{fig10}.
Our explanation for this maximum is heuristic.
The $\tau_{1}^{*}$ and $\tau_{2}^{*}$ parameters have been fitted to the dipolar energy or the diffusion constant.
Therefore, they are aggregated parameters that contain many effects averaged into them.
Judging from the maximum, we must have competing effects in play.

\begin{figure}[b]
	\centering
	\includegraphics[width=0.32\textwidth]{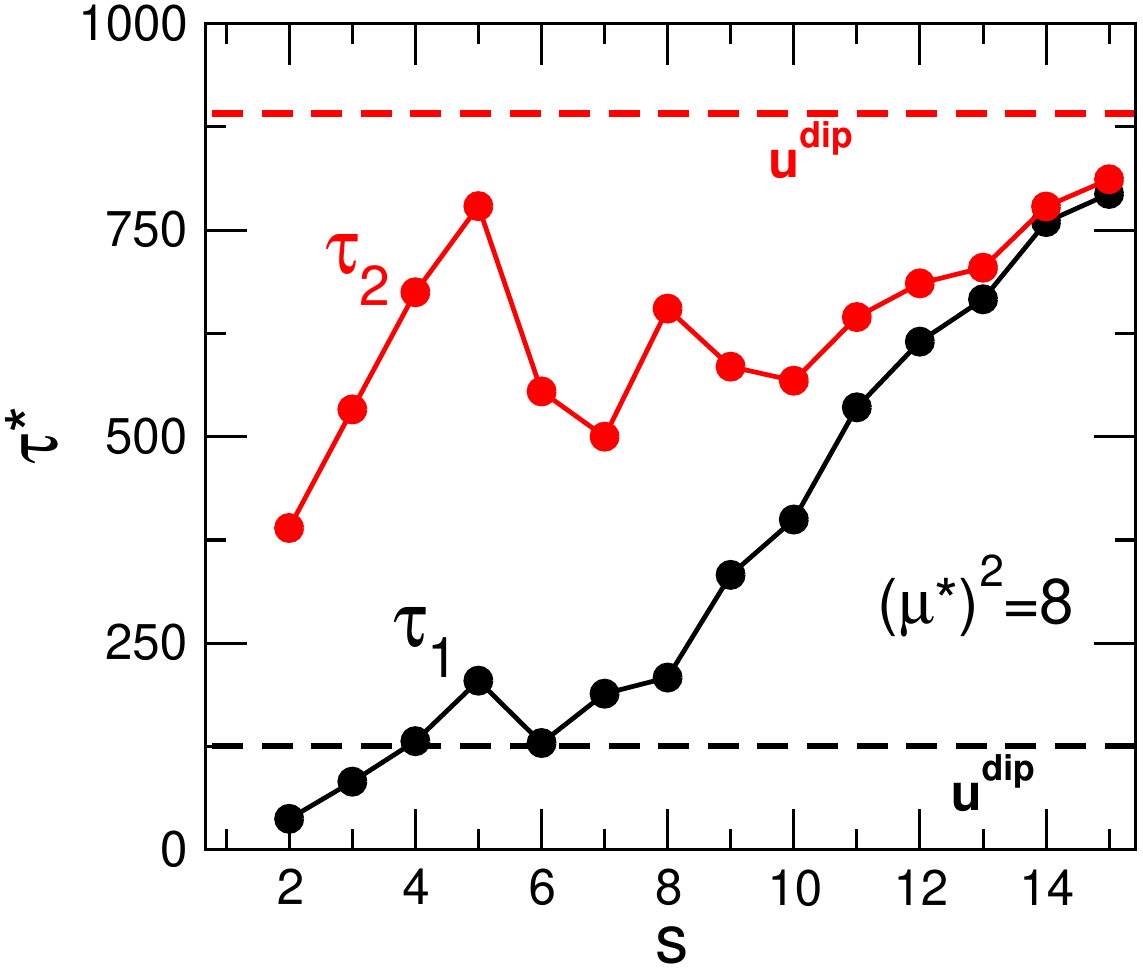}
	\caption{The $\tau^{*}_{1}$ and $\tau^{*}_{2}$ time constants obtained from bi-exponential fits on the $n_{s}(t^{*})$ functions as functions of $s$ for $(\mu^{*})^{2}{=}8$. The dashed lines represent the values obtained from fitting to the $(u^{\mathrm{dip}})^{*}(t^{*})$ function. Parameters are the same as in Fig.~\ref{fig07}.}
	\label{fig11}
\end{figure}

One effect is that larger time constants belong to the longer chains (see Fig.~\ref{fig11}).
With increasing $\mu^{*}$, longer chains appear in the system, slowing down the overall processes and resulting in larger time constants.
At the same time, stronger dipolar interactions speed up aggregation resulting in smaller time constants.
The two effects seem to compete resulting in the maximum.
At small values of $\mu^{*}$ (increasing part in Fig.~\ref{fig10}) the first effect dominates: longer chains just appear in the system.
When chain formation form is strong (above $(\mu^{*})^{2}{=}5$) the increasing dipolar attraction pulling these chains together is the dominating effect.

Fig.~\ref{fig11} provides a relation between ``aggregate'' time constants fitted to, for example, the dipolar energy, and time constants fitted to individual $n_{s}(t^{*})$ functions.
This figure shows the fitted time constants as functions of $s$ for $(\mu^{*})^{2}{=}8$.
The two horizontal lines indicate the time constants fitted to the dipolar energy.

We were able to fit to $n_{s}(t^{*})$ curves for not too large $s$ values.
Above $s{=}15$ fitting a bi-exponential does not really work, and the two time constants become equal.
It is a nice result that the ``aggregate'' values of $\tau_{1}^{*}$ and $\tau_{2}^{*}$ fitted to the energy confine the values fitted to the $n_{s}(t^{*})$ curves.

This result is in agreement with our hypothesis on the meaning of the two time constants.
The smaller time constant fitted to the energy is associated with the smaller time constants (formation) of the shorter chains.
The larger time constant fitted to the energy is associated with the time constants of the longer chains.
Because we start from monomers and dimers, there must be a fast process producing the intermediate products with time constant $\tau_{1}^{*}$.
When the chains are formed, however, they have their own dynamics with diffusion, dissociation, and association characterized by the time constant $\tau_{2}^{*}$

\subsection{The effect of packing}
\label{sec:rhodep}

The reduced density characterizes the packing of the particles; $\rho^{*}{=}Nd^{3}/V$ is the fraction of the volume occupied by the cubes around the particles in relation to the total volume. 
Note that an alternative reduced quantity is the packing fraction, $\rho^{*}\pi /6$, that is the fraction of the volume occupied by the particles themselves in relation to the total volume (it cannot be larger than $1$).
At larger $\rho^{*}$, the particles get close to each other to contact position ($r{=}d$) with a higher probability.

\begin{figure}[t]
	\centering
	\includegraphics[width=0.48\textwidth]{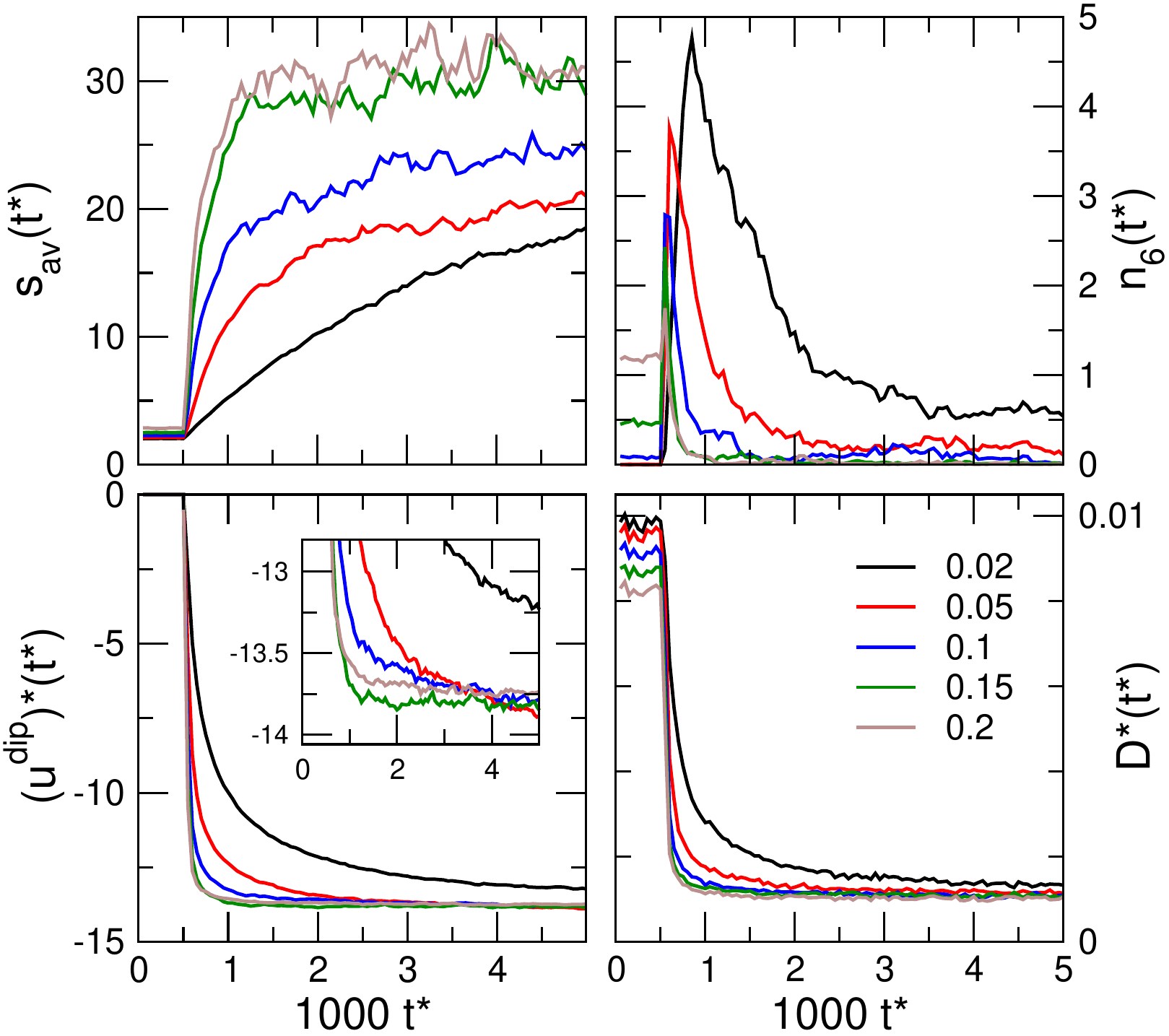}
	\caption{Analysis of the dependence of average chain length (top-left panel), number of chains of length $6$ (top-right panel), one particle dipolar energy (bottom-left panel), and diffusion constant (bottom-right panel) on the reduced density (packing).   Simulation parameters are $N{=}256$, $(\mu^{*})^{2}{=}6$, and $\gamma^{*}{=}100$.}
	\label{fig12}
\end{figure}

Fig.~\ref{fig12} shows the time dependence of various quantities for different values of $\rho^{*}$.
Interestingly, the one-particle dipolar energy goes to the same equilibrium value at different densities (bottom-left panel).
This is in contrast to the behavior of homogeneous isotropic bulk dipolar fluids (the dipoles can rotate there), where this energy sensitively depends on $\rho^{*}$.
The fact that in the case of the ER fluid with dipoles aligned in the $z$ direction $(u^{\mathrm{dip}})^{*}$ does not depend on $\rho^{*}$ implies that the dominant effect determining the dipole-dipole energy is the interactions of particles inside the chains.

At larger $\rho^{*}$ values, however, the curves approach the equilibrium values faster.
Stronger packing rather has effects on dynamics, because the particles can find each other faster.

\begin{figure}[b]
	\centering
	\includegraphics[width=0.3\textwidth]{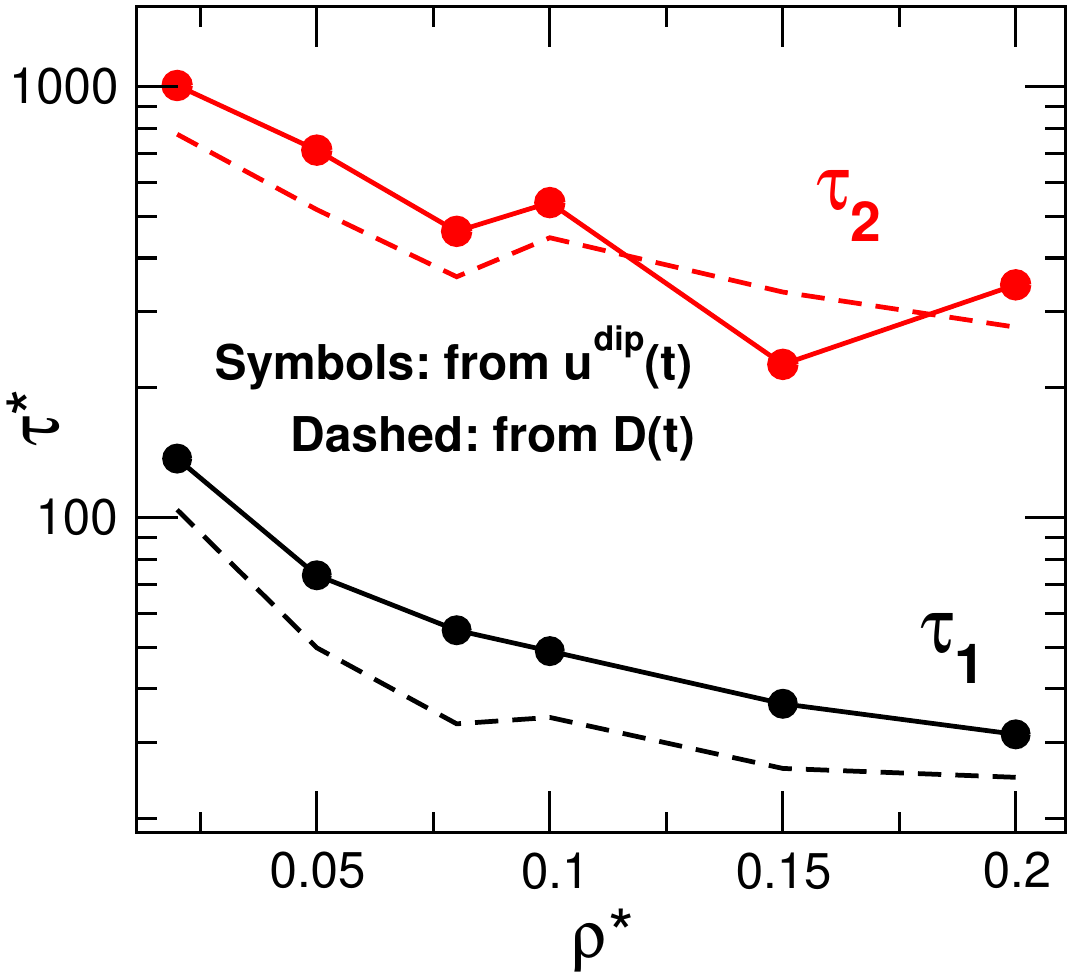}
	\caption{The $\tau_{1}$ and $\tau_{2}$ time constants obtained from bi-exponential fits to $(u^{\mathrm{dip}})^{*}(t^{*})$ (symbols) and $D^{*}(t^{*})$ (dashed lines) functions plotted against the reduced density, $\rho^{*}$. Parameters are the same as in Fig.~\ref{fig12}.}
	\label{fig13}
\end{figure}

What was said for the dipolar energy above is also valid for the diffusion constant (bottom-right panel).
Fig.~\ref{fig13} shows that the characteristic times, $\tau_{1}^{*}$ and $\tau_{2}^{*}$, get smaller as $\rho^{*}$ increases.
The $\tau_{1}^{*}$ and $\tau_{2}^{*}$ values fitted to $(u^{\mathrm{dip}})^{*}$ and $D^{*}$ behave similarly.
Although the uncertainty of fitting for $\tau_{2}^{*}$ is quite large, from this and earlier figures (Figs.~\ref{fig06}, \ref{fig10}, and \ref{fig11}) we can conclude that the time constant $\tau_{2}^{*}$ is larger than $\tau_{1}^{*}$ with about an order of magnitude. 

The top panels of Fig.~\ref{fig12} show the behavior of the average chain length, $s_{\mathrm{av}}(t^{*})$, and $n_{6}(t^{*})$ as a function of time.
The $s{=}6$ chains form and vanish more quickly at high densities that is also reflected in the time constants (Fig.~\ref{fig13}).

\begin{figure}[t]
	\centering
	\includegraphics[width=0.49\textwidth]{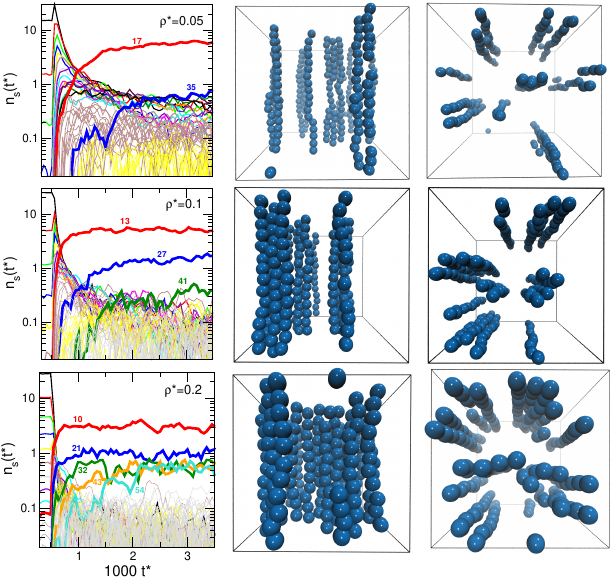}
	\caption{The number of chains with varying lengths ($s\geq 2$) averaged over blocks as a function of time, $t^{*}$ (first column) for three reduced densities ($\rho^{*}=0.05$, $0.1$, and $0.2$.), and snapshot from the simulation for each density at $t^{*}=50$, front (second column) and top views (third columns). Parameters are the same as in Fig.~\ref{fig12}.}
	\label{fig14}
\end{figure}

The average chain length increases with increasing $\rho^{*}$ despite the fact that the length of the chain overarching the cell decreases ($s_{0}{=}17$, $13$, and $10$ for the reduced densities $\rho^{*}{=}0.05$, $0.1$, and $0.2$, respectively, indicated by thick red lines in Fig.~\ref{fig14}).
This can only be caused by the fact that the chains aggregate at larger densities with higher probability.

This can be followed in more detail in Fig.~\ref{fig14} that shows the $n_{s}(t^{*})$ functions (similar to Fig.~\ref{fig09}) for three different densities (from top to bottom) with snapshots taken at $t^{*}{=}50$ when the chains are already formed.
The rows of the figure refer to reduced densities $\rho^{*}{=}0.05$, $0.1$, and $0.2$ (from top to bottom).

In the case of $\rho^{*}{=}0.02$, there was only one overarching chain (with length $s_{0}{=}23$, top-middle panel of Fig.~\ref{fig09} is the closest case) ``jumping out''.
At $\rho^{*}{=}0.05$, we have two: the single overarching chain ($s{=}17$) and two of them stuck together ($s{=}35$).
The double chain here and in the following rows are indicated with thick blue lines.
As the reduced density increases, triplets (green thick lines) and quadruplets (orange thick line) of chains appear.

The $n_{s}(t^{*})$ curves for $s<s_{0}$ are plotted with thin lines of various colors.
The $n_{s}(t^{*})$ curves between $s_{0}$ and lengths of chain pairs ($35$, $27$, $21$) are plotted with thin brown lines.
The $n_{s}(t^{*})$ curves between the pairs and triplets are plotted with thin yellow lines.
The $n_{s}(t^{*})$ curves above the triplet lengths are plotted with thin gray lines.
As $\rho^{*}$ increases, the number of long chains increases.
In the case of $\rho^{*}{=}0.2$, the gray curves dominate (next to those that ``jump out'').
This indicates that, beyond the overarching chains and their pairs, triplets, and so on, the dominant configurations are shorter chains attached to triplets, quadruplets, and so on.

\section{Conclusions and future prospects}

Our Brownian dynamics simulations revealed several interesting phenomena regarding the dynamics of chain formation in ER fluids.
Here, we focused on the approximate model where the particle-particle polarization is ignored (in accordance with the literature of similar studies), but preliminary results indicate that particle-particle polarization is important that should be taken into account.
The results obtained from this approximation should be considered as a qualitative description of the phenomena.
Particle-particle polarization will expectedly nuance the picture.

The parameter space that we considered corresponds to cases where the system is fluid-like (not frozen), quickly evolving (small friction coefficient), and relatively small ($N{=}256$).
All these served our purpose of performing a large bulk of simulations and provide a systematic study of the behavior of the system in dependence of the various parameters.
We did our best, however, in providing insight into how our results can be extrapolated to parameters outside the parameter space simulated here.
In future studies, however, those parameter sets can be considered if they prove to  be related to specific experimental situations under focus.

The behavior of the system under stress is a case that is crucially relevant from a technological point of view. 
This is also on our list.
Obviously, there is a lot to do in order to understand the microscopic level behavior of the ER devices.

\section*{Acknowledgments}

This research was supported by the European Union, co-financed by the European Social Fund, EFOP-3.6.2-16-2017-00002.
We also acknowledge the support of the National Research, Development and Innovation Office (NKFIH), project No.~K124353. 
We would like to thank Zolt\'an Hat\'o for his help in creating the figures and Tam\'as Varga for helpful discussions.

\end{document}